\documentclass[manuscript]{aastex}

\usepackage{color}

\slugcomment{Draft (\today) to be submitted to ApJ Supplement}

\shorttitle{Spectroscopy of Galactic Globular Clusters}
\shortauthors{Kim et al.}

\begin{document}

\title{A NEW CATALOG OF HOMOGENISED ABSORPTION LINE INDICES FOR MILKY WAY GLOBULAR CLUSTERS FROM HIGH-RESOLUTION INTEGRATED SPECTROSCOPY}

\author
  {Hak-Sub~Kim\altaffilmark{1,2},
  Jaeil~Cho\altaffilmark{1,3},
  Ray M.~Sharples\altaffilmark{4},
  Alexandre~Vazdekis\altaffilmark{5,6}, 
  Michael~A.~Beasley\altaffilmark{5,6} and
  Suk-Jin~Yoon\altaffilmark{2,7}
   }

\affil{
\altaffilmark{1}Co-first authors\\
\altaffilmark{2}Center for Galaxy Evolution Research, Yonsei University, Seoul 03722, Republic of Korea\\
\altaffilmark{3}Gwacheon National Science Museum, Gyeonggi 13817, Republic of Korea\\
\altaffilmark{4}Department of Physics, University of Durham, South Road, Durham DH1 3LE, UK\\
\altaffilmark{5}Instituto de Astrofsica de Canarias, La Laguna, E-38200 Tenerife, Spain\\
\altaffilmark{6}Departamento de Astrofsica, Univrsidad de La Laguna, Spain\\
\altaffilmark{7}Department of Astronomy, Yonsei University, Seoul 03722, Republic of Korea\\
\email{sjyoon0691@yonsei.ac.kr}}

\begin{abstract}
We perform integrated spectroscopy of 24 Galactic globular clusters. 
Spectra are observed from one core radius for each cluster
with a high wavelength resolution of ~$\sim$\,2.0\,\AA~FWHM.
In combination with two existing data sets from \citet{puz02} and \citet{sch05},  
we construct a large database of Lick spectral indices
for a total of 53 Galactic globular clusters
with a wide range of metallicities, $\rm-2.4 \la [Fe/H] \la 0.1$, 
and various horizontal-branch morphologies.
The empirical index-to-metallicity conversion relationships are provided for the 20 Lick indices 
for the use of deriving metallicities for remote, unresolved stellar systems.
\end{abstract}


\section{INTRODUCTION}

Globular clusters (GCs) are thought to have formed along with the bulk of stars in galaxies
and thus contain crucial information on the formation histories of their host galaxies.
The properties of GCs in external galaxies are derived by comparing their integrated light with Galactic globular clusters (GGCs).
Furthermore, reliable databases for GGCs are crucial for validating and calibrating theoretical stellar population models.
Many studies (e.g., \citealt{bur84}; \citealt{cov95}; \citealt{coh98}; \citealt{tra98}; \citealt{puz02}, hereafter PSK02; \citealt{sch05},
hereafter SRC05; \citealt{sch12}, hereafter S12; \citealt{pip11}; \citealt{roe14}) have investigated the line-strength indices from integrated spectra of GGCs. Among them, PSK02 presents a set of Lick indices for 12 GGCs including metal-rich bulge GCs.
SRC05 provides the largest set of integrated spectra for 41 GGCs (including most of the PSK02 sample), and S12 presents the Lick index measurements for the spectra later on. 

The Lick index system (\citealt{bur84}; \citealt{wor94}; \citealt{wor97}; \citealt{tra98}) is the most widely used spectral index system, which consists 25 line indices in the optical wavelength range (4000\,\AA~$\la~\lambda~\la$~6400\,\AA). This system has been renewed and
upgraded by several authors (e.g. \citealt{sch07}, \citealt{fra10}, \citealt{vaz10}) along with the improvement of modern instruments.
In this paper, we present 20 absorption line indices measured in high-resolution integrated spectra of 24 GGCs, 13 of which are newly observed. The line indices are calibrated to the Lick indices both on the \citet{sch07} redefinition (hereafter S07) and on the line-index system (hereafter LIS; \citealt{vaz10}). Our goal is to provide a combined catalog of widely used line-indices for the largest GGC sample.

The paper is organized as follows: 
Sections 2 and 3 describe the observation and data reduction, respectively.
In Section 4, we construct a homogeneous data set of the Lick indices on the S07 and LIS systems 
for 53 GGCs by combining our data with the existing catalogs, 
and derive empirical index-to-metallicity conversion relations 
for the Lick indices from the combined catalog.
We provide the fully reduced, flux-calibrated GC spectra as well.    
Section 5 summarizes our results.

\section{OBSERVATION}
 
Spectroscopic observations of 24 GGCs were conducted
with the 2.5-m Isaac Newton Telescope in La Palma, Spain from July 4--7, 2000.
The nights were not photometric.
Our sample of GGCs spans a wide range of metallicities,
$\rm-2.4 \la [Fe/H] \la 0.1$, with a mean metallicity of $\rm[Fe/H] \simeq -1.3$, on the \citet{Carr97} scale.
The sample also includes both clusters with blue horizontal branches and red horizontal branches.
The basic properties of these GGCs are listed in Table \ref{tab:mwgc}.
We used the Intermediate Dispersion Spectrograph (IDS) with the 235 camera,
the EEV10 CCD detector, the R900V grating, and a long-slit with a width of 1\farcs5.
Because of the severe vignetting of the 235 camera optics\footnote{See http://www.ing.iac.es/astronomy/instruments/ids/ids\_eev10.html},
the maximum useful slit length is 3.3$\arcmin$.
This configuration provides a spatial resolution of 0.4 arcsec~pixel$^{-1}$, a wavelength range of 4000\,--\,5400\,\AA, 
a spectral dispersion of 0.63\,\AA~pixel$^{-1}$, and a spectral resolution of FWHM~$\sim$\,2.0\,\AA. 
The FWHM values measured from arc lines are ~$\sim$\,1.94\,\AA~and do not show any trend with the wavelength.
The slit was drifted $\pm1\,r_c$ from the cluster center, where $r_c$ is the cluster core radius,
during an exposure time of 900 seconds.
In order to check the self-consistency of our data and foreground star contamination,
we repeated exposures with the slit positioned in an orthogonal direction.

\begin{figure}
\begin{center}
\includegraphics[scale=0.7]{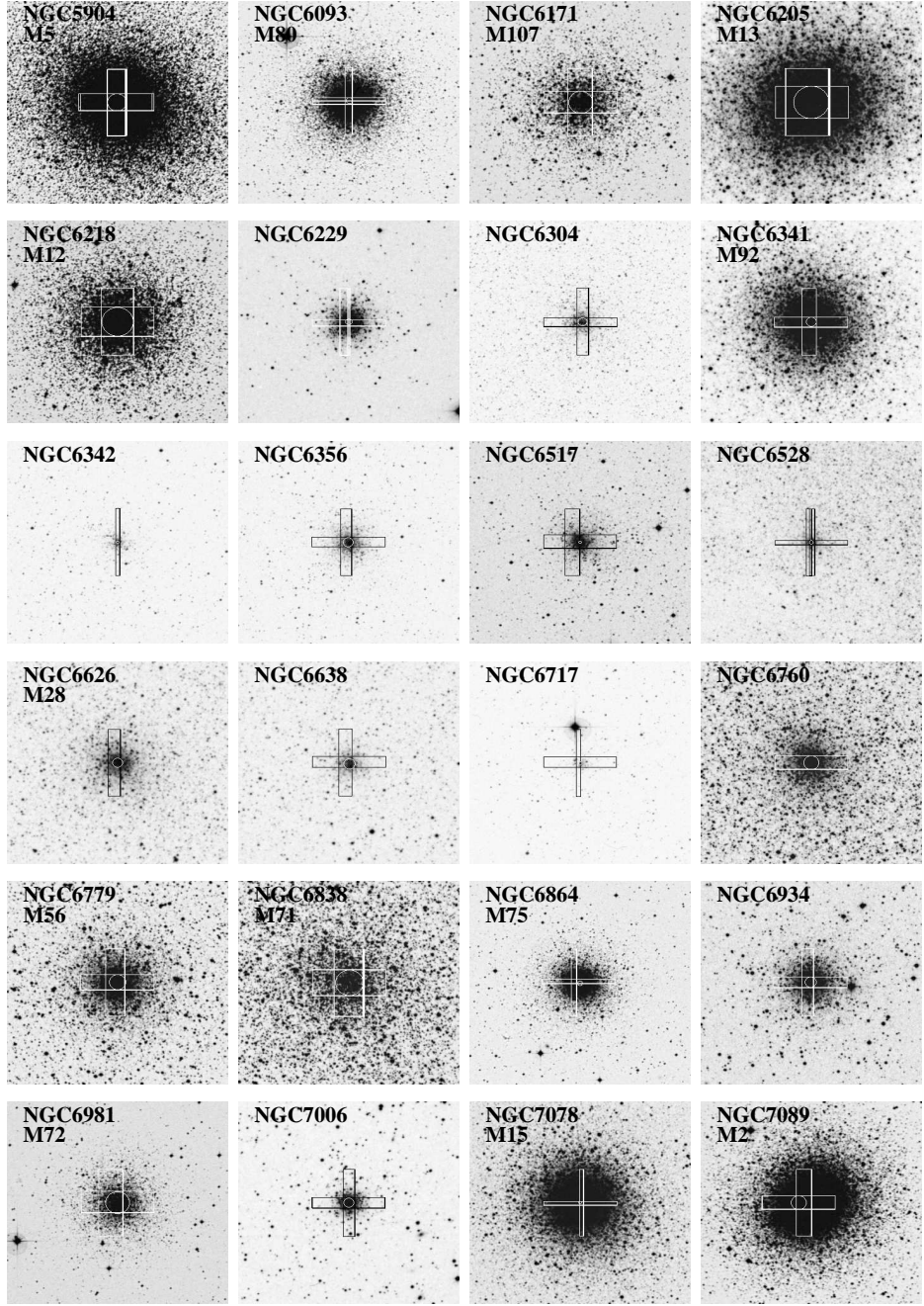}
\caption{Digitalized Sky Survey images of the GGCs observed along with the slit coverages. The circle in the center of each GC represents the core radius. In each rectangle, the long dimension is the slit size, and the short dimension is the drift coverages determined to cover a core radius from the center. The slit position and covering area are drawn based on the telescope positions in image headers from our observations and may have some pointing uncertainties.  }
  \label{fig:mwgcview}
 \end{center} 
\end{figure}

For large GCs, separate sky exposures (15$\arcmin$ north for NGC 5904 and NGC 6205, and 15$\arcmin$ north and 15$\arcmin$ east for NGC 6838) were obtained for background subtraction.  
The observation date, the number of exposures in each direction, the number of sky exposures, 
and the extraction windows are listed in Table \ref{tab:mwgc_log}.  
Figure \ref{fig:mwgcview} demonstrates the scan coverage for each GGC on Digitalized Sky Survey images. 
A total of 47 Lick standard stars were observed for accurate calibration to the standard Lick system. 
A spectrophotometric standard star was also observed each night for flux calibration. 
The exposure times for the standards varied from 4 seconds to 10 seconds depending on brightness. 
Between each repositioning of the telescope, an arc frame was taken using a CuAr/CuHe lamp for wavelength calibration. 
Each exposure produced a raw image with 500 pixels along the spatial axis and 4200 pixels along the dispersion axis. 
The CCD gain was $1.17\,e^{-}/\rm ADU$, and the readout noise was $4.2\,e^{-}/\rm pixel$.

\section{DATA REDUCTION}
\label{sec:redu}

The raw images were reduced using the standard IRAF package. 
The basic CCD reduction was performed according to the procedure 
described in {\it A User's Guide to CCD Reductions with IRAF} \citep{mas97}. 
The {\it IRAF/ccdred} package was first used to trim an overscan region and to subtract a bias frame. 
A flat-field frame and a twilight sky frame for each night were constructed by averaging several separate frames. 
Then, a normalized flat-field was obtained along the dispersion axis by fitting a fifth-order polynomial function using the {\it RESPONSE} task in the {\it twodspec.longslit} package, and a smoothed twilight sky flat was made along the spatial axis using the {\it ILLUM} task in the same package.
By multiplying the previous two products, the `ideal' flat was produced and
used for flat fielding of all scientific data as well as the calibration data. 
Cosmic rays were removed using the {\it APALL} routine in the IRAF package, 
which rejects any highly deviated values while optimally extracting a spectrum. 

After the basic CCD reduction, spectra were extracted and calibrated 
following the procedures in {\it A User's Guide to Reducing Slit Spectra with IRAF} \citep{mas92}. 
The {\it APALL} routine was used to extract a spectrum from the two-dimensional long-slit spectroscopic images. 
We set aperture radii for spectra extraction initially as $0.5\,r_c$ or $1.0\,r_c$ for each GGC, and increased to cover the outer region of the cluster.
For NGC 6342, the $0.5\,r_c$ extraction window was not used because of the small size of the cluster.
For the GGCs that did not have separate sky exposures, 
sky background regions were defined on both sides 
well away from the central light profile, avoiding any bright field stars. 
If one side of the outskirts of a GGC was heavily contaminated by field stars, 
the other side was selected to estimate the sky background. 
Figure \ref{fig:gc_apr} shows the flux summation along the dispersion direction, 
visualizing the spectrum extraction regions of the GGCs and the sky background areas. 
The extraction regions were selected to cover almost the whole area in the spatial axis avoiding any bright stars
for images with separate sky exposures. 
Once these extraction windows for the GGCs and sky backgrounds were defined, 
the {\it APALL} routine automatically determined a center for the GGCs in the spatial axis 
and traced it along the dispersion axis with a ten pixel step size. 
The process then combined the spectra within the previously defined extraction windows and subtracted the sky spectrum. 
During this process, cosmic rays were rejected according to three-sigma clipping. 
For images with independent sky exposures (see Table \ref{tab:mwgc_log}), the sky backgrounds were subtracted manually 
at the 1D extracted spectrum level.
Most results in this work are derived from spectra with an aperture size of $1\,r_c$.
For standard star calibration, a fixed aperture radius was set at five pixels with a background region 
between 10 and 20 pixels on each side of the object. 

Using CuAr lines in the arc frames for each target, 
the scientific and calibration data were calibrated with an $\rm r.m.s.$\,$\sim$\,0.2\,\AA~precision. 
Flux calibration was done with the spectrophotometric standard stars Feige110, BD+33-2642, and BD+26-2606 (\citealt{oke90}; \citealt{mas88}). 
The instrumental sensitivity function was determined using the standard stars from all nights. We verified that the continua of the GC spectra, when calibrated with the sensitivity function and corrected for reddening 
using the $E(B-V)$ values from \citet{har96},
show generally good agreements with those of model spectra of the same ages and metallicities with observed GCs predicted from the Yonsei Evolutionary Populations Synthesis (YEPS) model \citep{Chung13}.

\section{RESULTS AND DISCUSSION}

\subsection{Integrated Spectra}
\label{sec:library}

Figure~\ref{fig:spec_gcs} presents the reduced integrated spectra of the GGCs observed in this study. The wavelength- and flux-calibrated GGC spectra are corrected for Galactic extinction and shifted to the rest-frame. We combine all the spectra of each GGC and shifted each spectrum vertically by an arbitrary amount for clarity. The spectra are ordered by increasing metallicity. The GC name with metallicity, horizontal-branch ratio, and signal-to-noise ratio around 4700\,\AA~are denoted above each spectrum. The signal-to-noise ratio per pixel are calculated using the IDL function {\it DER\_SNR} \citep{sto08}. NGC 6760 is excluded from the figure and further analysis due to its low signal-to-noise ratio.

Comparing the spectra from the two orthogonal slit directions allows to check the self-consistency of our data and any possible field star contamination. Both the drift coverage and the extraction windows are $\pm1\,r_c$ from the center, so that the spectra from the two scan directions cover the same region of the cluster. Hence the two spectra should be identical in principle, despite the different scan directions. This overlapping area is defined by the two orthogonal exposures as shown in Figure \ref{fig:mwgcview}. 
The comparison shows that most GCs show good agreement between the spectra obtained from different scan directions.
On the other hand, some GCs show differences in spectral shape even between the spectra of the same scan direction. 
In Figure~\ref{fig:spec_gcs2}, we present examples of GC spectra showing variances in the continuum flux level (upper panel) and/or the spectral slope (lower panel) between individual exposures. NGC 6779, NGC 6218, NGC 6717, NGC 6342, and NGC 6304 show flux differences, and NGC 6093 and NGC 7078 show discernible differences in the spectral slope.
These disagreements may be due to the telescope pointing errors and the variations in sky conditions during our observations, 
and to the differences in the sky spectra for the cases where the sky spectra were taken from the edge of the slit. 
All of these spectral variations would cause uncertainty in spectral indices, and the uncertainty is reflected in the index error estimation as described in Section~\ref{sec:Index}.

We provide the wavelength- and flux-calibrated GGC spectra extracted from a series of apertures listed in Table~\ref{tab:mwgc_log}.
The data are in the multispec FITS format comprised of four bands.
The first band contains variance-weighted, cosmic-ray cleaned, background-subtracted spectrum. The second band contains background-subtracted spectrum without variance-weighting and cosmic-ray cleaning. The third band contains background spectrum obtained from a separate sky exposure or the ends of the long slit as mentioned in Section~\ref{sec:redu}. The fourth band contains the sigma spectrum.
The spectra are made available at the YEPS\footnote{http://web.yonsei.ac.kr/cosmic} and the MILES\footnote{http://miles.iac.es} websites.

\subsection{Index Measurements}
\label{sec:Index}

\subsubsection{Radial Velocities}
\label{sec:RadialV}

Radial velocities are measured by the penalized pixel-fitting method (pPXF; \citealt{cap04}).
The method extracts the line-of-sight velocity distributions (LOSVD) described by the Gauss-Hermite series \citep{ger93, van93} by directly fitting observed spectra to template spectra in pixel space. 
In order to avoid the template-mismatch problem, we use 350 MILES models \citep{vaz10} 
as velocity templates covering a wide range of ages (0.06\,--\,17 Gyr) and 
metallicities (--2.3 $\leq$ [m/H] $\leq$ 0.2). The MILES models have 
a slightly lower spectral resolution than our data (2.51\,\AA~versus 
2.0\,\AA~FWHM) and we smooth our spectra to the MILES resolution.
Following the procedure described in \citet{cap04}, we determine the optimized penalty parameter for each GGC spectrum which minimizes the uncertainties in fitting with the high order Gauss-Hermite moments (h3 and h4) 
by biasing the fitting solution towards a Gaussian LOSVD. 
We then derive the radial velocities by running pPXF on each scan of the target GGCs in the wavelength range of 4000\,\AA~to~5400\,\AA. The velocity uncertainties are determined as the standard deviation of the velocities measured from 500 Monte Carlo simulations for each GGC spectrum, in which the velocity measurements are repeated with the template spectra slightly modified by adding noise. 

To determine the final velocity of each GGC, we compare the velocities measured from different scan directions for the same object. We found that, in some cases, there are significant velocity differences between scan directions, which seems to be originated from the zero point shift in wavelength calibrations. We have taken the arc frames between the repositioning of the telescope but not between the change of the slit direction. This leads to wavelength zero point shift for some objects resulting in the velocity zero-point offsets between the data. We therefore apply zero point corrections to the velocity measurement using the [O I] $\lambda$5577 night sky line. The final velocity of each GGC is determined as the error-weighted mean of the velocities measured from different scans and the velocity error is computed using standard error propagation procedures.

We note that the typical velocity dispersion of GGCs ($<$\,10\,km/s; \citealt{dub97}) is smaller than our instrumental velocity resolution ($\sigma\thicksim$ 54.2~km s$^{-1}$). 
The pPXF method have had an issue that the kinematics could not be reliably recovered when the velocity dispersion of the object 
is smaller than the instrumental dispersion. However, the pPXF code has recently been updated solving the issue by adopting the analytic Fourier transform of the LOSVD (for details, see Section 4 of \citealt{cap16}). 
To check the robustness of our measurements, we also measure the radial velocities by Fourier cross-correlation using the {\it FXCOR} task in IRAF. The measurements are generally in good agreement with each other, whereas the velocity measurement via pPXF method gives more consistent values for the same GGC.

Table \ref{tab:mwgc_rd} lists the final radial velocities and their uncertainties along with the radial velocities from \citet{har96}. 
Figure~\ref{fig:com_gcvel} compares our measurements and Harris's values. 
Our velocities are in generally good agreement with the literature values.

\subsubsection{Measuring the Lick Indices on the S07 System}
\label{sec:Lick}

S07 proposed a new Lick index system (referred to as S07 system in this paper) which is based on the \citet{jon99} spectral library containing a large set of high-resolution, flux-calibrated stellar spectra. By adopting the flux-calibrated standard spectra, the S07 system is free from the possible uncertainties associated with the response curve of the original Lick/IDS spectrograph and achieves a higher accuracy of the Lick indices as illustrated in Figures 1 and 2 of S07.

We measure the equivalent widths (EWs) of the indices for all individual GGC spectra, following the definition of the index passbands provided by \citet{wor94} and \citet{wor97}, which consists 25 line indices in the optical wavelength range (4000\,\AA~$\la~\lambda~\la$~6500\,\AA). Our measurements are restricted to 20 indices excluding the five longest wavelength ones (Fe5709, Fe5782, NaD, TiO${_1}$, and TiO${_2}$) because of the limited wavelength coverage (4000\,\AA\,$\la\,\lambda\,\la$\,5400\,\AA) of our spectra.
Before measuring the indices, we shift our GGC spectra to the rest-frame using the radial velocities determined in Section~\ref{sec:RadialV}, and degrade the spectra to the wavelength-dependent resolution of the Lick index system\footnote{The resolutions are given in the Table 1 of S07. The S07 system is defined on the original Lick/IDS variable resolution because the index measurements of \citet{wor94} 
were used as a supplement to the database.}. 
We then measure the EWs of the indices and index errors from the spectra with the {\it LICK\_EW} code in the EZ\_Ages IDL package \citep{gra08}. The line broadening caused by velocity dispersion is not considered since GCs, unlike galaxies, have very low velocity dispersion ($<$\,10\,km/s; \citealt{dub97}). 

We calibrate the instrumental index measurements to the S07 system using the Lick standard stars. 
The index measurements for the standard stars are performed in the same way as for the GGCs. 
In Figure \ref{fig:lick_cal}, our index measurements for the standard stars are compared with those reported in S07\footnote{Some indices are not provided by S07 due to gaps in the coverage of the Jones (1999) spectra. For those measurements, we use the Indo-US stellar spectra that are in common with the original Lick/IDS library. We obtained the data via private communication with Schiavon.}.
The differences between the two measurements are generally small compared to the observational error and only minor zero-point offset corrections are needed. The zero-point offsets are determined by calculating the error-weighted mean of the differences after removing outliers by applying a three-sigma clipping procedure.
The resulting offsets, $\Delta_{index}=EW_{S07}-EW_{this~work}$, are listed in Table \ref{tab:lick_off}. 
The instrumental indices are converted to the Lick/S07 system by adding the offset values. The final Lick/S07 indices for each GGC are calculated by the error-weighted mean of the Lick/S07-calibrated indices from multiple exposures, and the index errors are determined as the standard deviation of the error-weighted mean.

\subsubsection{Measuring the Lick Indices on the LIS System}
\label{sec:LIS}

\citet{vaz10} proposed the LIS system as an alternative to the Lick/IDS system
aiming to minimize the uncertainties
caused by degrading the observed spectral resolution to the variable resolution of the Lick/IDS definition (8--11\,\AA).
The LIS system is defined on a flux-calibrated spectrum whose spectral resolution is constant along the whole spectral range.
The system uses three standard resolutions of 5, 8.4, and 14\,\AA~FWHM, which are suitable for studies of GCs, intermediate-mass galaxies, and massive galaxies, respectively. In this study, we use the same index definition of the Lick system on the 5.0\,\AA~LIS system.

To measure the Lick indices on the LIS system from our data, 
we smooth our GC spectra with a gaussian using the {\it GAUSS} task in IRAF with a sigma of 2.9 pixels to give FWHM\,$\sim$\,5.0 \AA.
We then shift the spectra to the rest-frame and measure the indices with {\it LECTOR}\footnote{http://www.iac.es/galeria/vazdekis/vazdekis\_software.html} software.
The final index values are determined by taking the error-weighted mean of the indices 
and the uncertainties are estimated by calculating the standard deviation of the error-weighted mean.

Figure~\ref{fig:lis_cal} shows a comparison between the index measurements for the standard stars in our data and for the stars provided in the MILES stellar library \citep{vaz10}. The indices on the LIS system, in principle, do not require any calibrations on a standard system if the measurements are based on well flux-calibrated spectra. However, as shown in the figure, there are some offsets which is likely due to the imperfectness of flux-calibrations. Hence, we decide to calibrate our measurements on the LIS system of \citet{vaz10} by adding the offsets to our measurements. The offsets are given in Table~\ref{tab:lis_off}.

\subsection{Data Compilation and the Combined Catalogs}
\label{sec:compre}

\subsubsection{The Catalog of Lick/S07 Indices}

We combine our results with previous catalogs of GGC Lick indices provided by PSK02 and by S12. PSK02 observed 12 GGCs mainly associated with the Galactic bulge using the ESO 1.52-m telescope in La Silla and measured all 25 Lick indices. SRC05 obtained integrated spectra of 41 GGCs with the Cerro Tololo Inter-American Observatory (CTIO) Blanco 4-m telescope and released their calibrated spectra with FWHM~$\sim$\,3.1\,\AA. S12 provides 23 Lick indices---except Fe4531 and Fe5015 due to the CCD defects in their spectral regions \citep{sch05, men07}---measured from the SRC05 spectra, which are calibrated to the S07 system. There are five common GCs between our data and PSK02, 11 common GCs between our data and S12, and four common GCs in all three sources. 

In Figure~\ref{fig:com_pu}, we compare our index measurements with those of PSK02 and S12 for the common GCs.
The systematic difference between our data and PSK02 data arises from the use of different Lick systems: PSK02 use the original Lick/IDS system with the index definition of \citet{wor97} and \citet{tra98} while our data are calibrated to the Lick/S07 system with the index definition of \citet{wor94} and \citet{wor97}. We determine the index offsets, $\Delta_{PSK02}=EW_{this~work}-EW_{PSK02}$, by computing error-weighted mean values of the differences between the two indices, and convert the PSK02 indices to the Lick/S07 system. The offset values are given in Table \ref{tab:gc_off}. On the other hand, the difference between ours and S12 data can be explained by random errors and/or the difference of the observed regions for a given GC.

The three datasets of Lick/S07 indices are merged to form a new one. For the common GCs, we adopt the error-weighted mean as the final index value. Table~\ref{tab:Lick} provides the combined catalog of 20 Lick/S07 indices for 53 GGCs covering a wide range of metallicities,  $\rm-2.4 \la [Fe/H] \la 0.1$. The index uncertainties listed under the Lick indices are the standard deviation of the error-weighted mean.

\subsubsection{The Catalog of LIS Indices} 

In addition to our spectral data, we use the integrated spectra of 41 GCs provided by SRC05 for constructing a final catalog of the LIS indices. We smooth the SRC05's spectra with FWHM~$\sim$\,3.1\,\AA using a gaussian of 1.67 pixel sigma to match the LIS-5.0 \AA~resolution. We then measure the LIS indices with {\it LECTOR} software from the spectra. For the multiple exposures of the same GC, the final index values are determined by taking the error-weighted mean of the index measurements and the uncertainties are estimated by calculating the standard deviation of the error-weighted mean.

The two datasets are combined to create the final LIS index catalog of a total of 53 GGCs. For the 11 common GCs in the two datasets, there are small zero point offsets which are likely due to the differences in the flux calibration between the two datasets as mentioned in Section~\ref{sec:LIS}. Before combining the two datasets, the zero point offset corrections are applied. For the Fe4531 and Fe5015 indices, we do not use the measurements of SRC05 data because of the known problems in their spectra. The index values adopted for the common GCs are the error-weighted mean of the two measurements and the uncertainties are determined by the standard deviation of the error-weighted mean. 
The final LIS index catalog is given in Table~\ref{tab:LIS}.

\subsection{The Empirical Index--Metallicity Relations} 

This Section provides empirical index--metallicity relations (IMRs) based on our combined index catalogs of 53 GGCs. One can use the IMRs to estimate the metallicities of extragalactic GCs. Determining the metallicities of extragalactic GCs is, however, not a straightforward task because of the well-known age--metallicity degeneracy, the effect of $\alpha$-elements enhancement, and even the ambiguity of the term ``metallicity" (e.g., \citealt{puz02, Beas08}). On the one hand, the metallicities of GGCs are relatively well constrained because they can be estimated from various ways including direct observations for the cluster member stars (e.g., \citealt{mal14}, \citealt{val15}, \citealt{mes15}).
Hence, with an assumption that extragalactic GCs are analogous to GGCs, calibrating the metallicity of extragalactic GCs to that of GGCs using the IMRs is a useful way to study the nature of extragalactic GCs  (e.g., \citealt{Bro90, coh98, Kis98, Nan10, pip11, Park12}). The IMRs are also useful for validating and calibrating theoretical stellar population models, as they provide observational constraints for comparison with model predictions (e.g., \citealt{Chung13, KimS13}).

We derive the empirical relations between both the Lick/S07 and the LIS indices and metallicity using our combined GGC catalog.
The metallicity values are taken from \citet{Carr09} who updated the scale of \citet{Carr97} metallicity scale.
Because both variables---spectral index and metallicity---have measurement errors, 
we adopt an orthogonal distance regression method \citep{Bogg90} to determine the best-fit polynomial function.
During the fitting procedure, outliers are rejected by applying the three-sigma clipping method.

Figure~\ref{fig:Indmet} shows the IMRs for 20 Lick/S07 indices. The blue, green, and red solid lines represent first-, second-, and third-order polynomial functions, respectively, i.e., 
\begin{equation}
\rm[Fe/H]\it=a_{0}+a_{1} \times (index),
\end{equation}
\begin{equation}
\rm[Fe/H]\it=a_{0}+a_{1} \times (index) + a_{2} \times (index)^2,\rm~and
\end{equation}
\vspace{-3.0mm}
\begin{equation}
\rm[Fe/H]\it=a_{0}+a_{1} \times (index) + a_{2} \times (index)^2 + a_{3} \times (index)^3.
\end{equation}

\hspace{-7.8mm}
The units of $\it index$ are mag for $\rm CN_{1},~CN_{2},~Mg_{1},~and~Mg_{2}$ and angstrom for the other indices. 
We also present the 95\% confidence bands of the LOESS regression, a nonparametric locally weighted regression \citep{clev94}, as gray-shaded regions for a visual comparison of the polynomial fits with the underlying trend of the data.
The polynomial coefficients are given in Table~\ref{tab:coeff1} (for the first order fit), Table~\ref{tab:coeff2} (second order), 
and Table~\ref{tab:coeff3} (third order). We also provide 
the Bayesian information criterion (BIC; \citealt{sch78}), which is a statistical criterion for model selection where 
the model with the lowest BIC would be the best model.
The BIC comparison between the polynomial fits show that the higher-order polynomials (green and red solid lines) give better fits than the first-order polynomial (blue solid line) for most indices,
which indicates the nonlinearity of the IMRs.
The nonlinearity is also implied by the LOESS results, particularly at the high and low metallicity ends.
The IMRs should be valid only in the range from the minimum to maximum values of each index,
which are listed in the last columns of Tables~\ref{tab:coeff1}--\ref{tab:coeff3}.

Figure~\ref{fig:LISmet} shows the IMRs for 20 LIS indices in the same format as in Figure~\ref{fig:Indmet}. 
Similar to Lick/S07 IMRs, the higher-order polynomials give better fit in general.
The polynomial coefficients are given in Table~\ref{tab:LIScoeff1} (first order), Table~\ref{tab:LIScoeff2} (second order), 
and Table~\ref{tab:LIScoeff3} (third order) along with the BIC.
The valid ranges of relations are given in the last columns of Tables~\ref{tab:LIScoeff1}--\ref{tab:LIScoeff3}.

A recent study by \citet{car11} and \citet{KimS13} showed with high precision spectroscopy of 280 GCs in M31
that strong inflection exists in the relations between metallicity and Balmer lines (H${\beta}$, H${\gamma_F}$, and H${\delta_F}$) and appreciable nonlinearity between metallicity and the Mg$b$ line.
Such nonlinear IMRs have been predicted by several stellar population simulation models which incorporate core helium burning horizontal-branch stars in GCs (e.g., \citealt{Lee05}; \citealt{Yoon06}; \citealt{Chung13}). 
In our catalog data, although not conclusive, the relationships between metallicity and Balmer lines appear inflected.
For Mg$b$, the nonlinear feature is weaker in our GGC data than that seen in the M31 GC data \citep{KimS13}.
This is probably due to the smaller sample size and the larger observational errors of our data
compared to \citet{car11} and \citet{KimS13} 
and/or to the larger intrinsic scatter in parameters such as age and ${\alpha}$-element mixture.
A detailed discussion on the nonlinearity issue is beyond the scope of this paper, and we refer the interested reader to \citet{Yoon06,Yoon11a,Yoon11b,Yoon13} for further discussion.

\section{SUMMARY}
\begin{itemize}

\item We obtained integrated spectra of 24 Galactic globular clusters with a high spectral resolution of FWHM~$\sim$\,2\,\AA~
using the 2.5-m Isaac Newton Telescope in La Palma, Spain. Our cluster sample spans a wide range of metallicities, $\rm-2.4 \la [Fe/H] \la 0.1$, and various horizontal-branch morphologies.

\item We measured 20 Lick indices in the wavelength range of $\rm 4000\,\AA\,\la\,\lambda\,\la\,5400\,\AA$ from the spectra. 
The Lick indices are calibrated both on the S07 and LIS systems, which are newly defined Lick index systems 
based on the modern stellar libraries.
We also measured the LIS indices from the spectra of 41 Galactic globular clusters provided 
by \citet{sch05} in the same manner.

\item For the largest Galactic globular cluster sample (53 clusters), we constructed combined catalogs of 20 Lick indices on the S07 
and LIS systems using our data and the data sets provided by \citet{puz02} and \citet{sch05}.

\item The combined catalogs and the fully reduced, flux-calibrated spectra can be found in the YEPS and MILES websites.

\item We derived the empirical index-to-metallicity conversion relations 
for both the Lick/S07 and the LIS indices, which can be used for extragalactic globular cluster studies.

\end{itemize}

\acknowledgments  
S.-J.Y. acknowledges support by
Mid-career Research Program (No. 2015-008049)
through the National Research Foundation (NRF) of Korea,
the NRF of Korea to the Center for Galaxy Evolution Research (No. 2010-0027910),
and the Yonsei University Future-leading Research Initiative of 2015-2016.
This work was supported in part by the Yonsei University Research Fund of 2013.
M.B. and A.V. acknowledge support from the Programa  Nacional de Astronom{\'{\i}}a y    
Astrof{\'{\i}}sica of MINECO,  under grant AYA2013-48226-C3-1-P.



\clearpage
\begin{deluxetable}{lcrrrrrr}
\tablecolumns{8}
\tablecaption{Sample GC properties}
\tablewidth{0pt}
\tablehead{\colhead{Name} & \colhead{Other Name} & \colhead{$l~\rm[^{\circ}]$} & \colhead{$b~\rm[^{\circ}]$} & \colhead{$R_{gc}~\rm[kpc]$} & \colhead{[Fe/H]$_{Harr}$} & \colhead{[Fe/H]$_{Carr}$} & \colhead{HBR} \\
\colhead{ } & \colhead{ } & \colhead{(1)} & \colhead{(2)} & \colhead{(3)} & \colhead{(4)} & \colhead{(5)} & \colhead{(6)}}
\startdata
 NGC~5904 &   M5 &   3.86 &  46.80 &  6.2 &  --1.29 &  --1.33 &  0.31 \\
 NGC~6093 &  M80 & 352.67 &  19.46 &  3.8 &  --1.75 &  --1.75 &  0.93 \\
 NGC~6171 & M107 &   3.37 &  23.01 &  3.3 &  --1.02 &  --1.03 &  --0.73 \\
 NGC~6205 &  M13 &  59.01 &  40.91 &  8.4 &  --1.53 &  --1.58 &  0.97 \\
 NGC~6218 &  M12 &  15.72 &  26.31 &  4.5 &  --1.37 &  --1.33 &  0.97 \\
 NGC~6229 &      &  73.64 &  40.31 & 29.8 &  --1.47 &  --1.43 &  0.24 \\
 NGC~6304 &      & 355.83 &   5.38 &  2.3 &  --0.45 &  --0.37 &  --1.00 \\
 NGC~6341 &  M92 &  68.34 &  34.86 &  9.6 &  --2.31 &  --2.35 &  0.91 \\
 NGC~6342 &      &   4.90 &   9.72 &  1.7 &  --0.55 &  --0.49 &  --1.00 \\
 NGC~6356 &      &   6.72 &  10.22 &  7.5 &  --0.40 &  --0.35 &  --1.00 \\
 NGC~6517 &      &  19.23 &   6.76 &  4.2 &  --1.23 &  --1.24 &   \nodata \\
 NGC~6528 &      &   1.14 &   --4.17 &  0.6 &  --0.11 &  0.07 &  --1.00 \\
 NGC~6626 &  M28 &   7.80 &   --5.58 &  2.7 &  --1.32 &  --1.46 &  0.90 \\
 NGC~6638 &      &   7.90 &   --7.15 &  2.2 &  --0.95 &  --0.99 &  --0.30 \\
 NGC~6717 & Pal~9 &  12.88 &  --10.90 &  2.4 &  --1.26 &  --1.26 &  0.98 \\
 NGC~6760 &      &  36.11 &   --3.92 &  4.8 &  --0.40 &  --0.40 &  --1.00 \\
 NGC~6779 &  M56 &  62.66 &   8.34 &  9.2 &  --1.98 &  --2.00 &  0.98 \\
 NGC~6838 &  M71 &  56.75 &   --4.56 &  6.7 &  --0.78 &  --0.82 &  --1.00 \\
 NGC~6864 &  M75 &  20.30 &  --25.75 & 14.7 &  --1.29 &  --1.29 &  --0.07 \\
 NGC~6934 &      &  52.10 &  --18.89 & 12.8 &  --1.47 &  --1.56 &  0.25 \\
 NGC~6981 &  M72 &  35.16 &  --32.68 & 12.9 &  --1.42 &  --1.48 &  0.14 \\
 NGC~7006 &      &  63.77 &  --19.41 & 38.5 &  --1.52 &  --1.46 &  --0.28 \\
 NGC~7078 &  M15 &  65.01 &  --27.31 & 10.4 &  --2.37 &  --2.33 &  0.67 \\
 NGC~7089 &   M2 &  53.37 &  --35.77 & 10.4 &  --1.65 &  --1.66 &  0.96 \\
\enddata
\vspace{ -8mm}
\tablecomments{(1) Galactic longitude; (2) Galactic latitude; (3) Distance from Galactic center; (4) Metallicity from \citet{har96}; (5) Metallicity from \citet{Carr09}; (6) Horizontal-branch ratio defined as $HBR\equiv (B-R)/(B+V+R)$ by \citet{lee94}, where B and R are the number of stars bluer and redder than the instability strip respectively and V is the number of RR Lyrae stars. The data are taken from \citet{har96} except [Fe/H]$_{Carr}$.}
\label{tab:mwgc}
\end{deluxetable}

\clearpage
\begin{deluxetable}{clcccc}
\tablecolumns{6}
\tablecaption{Observation log}
\tablewidth{0pt}
\tablehead{\colhead{Night} & \colhead{Object Name} & \colhead{S-N\tablenotemark{a}} & \colhead{W-E\tablenotemark{b}} & \colhead{Sky\tablenotemark{c}} & \colhead{Extraction Windows} \\
\colhead{} & \colhead{} & \colhead{} & \colhead{} & \colhead{} & \colhead{($r_c$)}}
\startdata
 4 Jul 2000 & NGC~5904~(M5) & 2 & 2 & 2 & 0.5, 1, 2, 3, 4\\
 	& NGC~6171~(M107) & 2 & 2 &   & 0.5, 1, 2, 3\\
	& NGC~6229	& 2 & 2 &   & 0.5, 1, 2, 3, 4, 5\\
	& NGC~6838~(M71) & 2 & 2 & 4 & 0.5, 1, 1.5, 2, 2.5\\
	& NGC~7089~(M2) & 2 & 2 &   & 0.5, 1, 2, 3, 4\\
 5 Jul 2000 & NGC~6205~(M13) & 2 & 2 & 1 & 0.5, 1, 1.5, 2\\
	& NGC~6528 & 2 & 2 &   & 0.5, 1, 2, 3, 4\\
	& NGC~6341~(M92) & 2 & 2 &   & 0.5, 1, 2, 3, 4\\
	& NGC~6934 & 2 & 2 &   & 0.5, 1, 2, 3, 4\\
	& NGC~7006 & 2 & 2 &   & 0.5, 1, 2, 3, 4\\
	& NGC~7078~(M15) & 2 & 2 &   & 0.5, 1, 2, 3, 5, 10\\
 6 Jul 2000 & NGC~6517 & 2 & 2 &   & 0.5, 1, 2, 3, 5, 10\\
	& NGC~6356 & 2 & 2 &   & 0.5, 1, 2, 3, 4\\
	& NGC~6638 & 2 & 2 &   & 0.5, 1, 2, 3, 4\\
	& NGC~6717 & 2 & 2 &   & 0.5, 1, 2, 3, 4\\
	& NGC~6864~(M75) & 2 & 2 &   & 0.5, 1, 2, 3, 4\\
	& NGC~6981~(M72) & 2 & 2 &   & 0.5, 1, 1.5, 2, 2.5\\
 7 Jul 2000 & NGC~6093~(M80) & 2 & 1 &   & 0.5, 1, 2, 3, 4, 5\\
	& NGC~6304 & 2 & 2 &   & 0.5, 1, 2, 3\\
	& NGC~6218~(M12) & 3 & 1 &   & 0.5, 1, 1.5, 2\\
	& NGC~6342 & 0 & 3 &   & 1, 2, 4, 6\\
	& NGC~6626~(M28) & 0 & 2 &   & 0.5, 1, 2, 3\\
	& NGC~6779~(M56) & 2 & 2 &   & 0.5, 1, 2, 3\\
	& NGC~6760 & 2 & 0 &   & 0.5, 1, 2, 3\\
\enddata
\vspace{-4mm}
\tablenotetext{a}{Number of exposures in the South-North scan direction}
\tablenotetext{b}{Number of exposures in the West-East scan direction}
\tablenotetext{c}{Separate sky exposures were taken for three bright extended globular clusters to obtain a better sky subtraction. In other cases, sky spectra were taken from the ends of the long slit.}
\label{tab:mwgc_log}
\end{deluxetable}

\clearpage

\clearpage
\begin{figure}
\begin{center}
\includegraphics[scale=0.95]{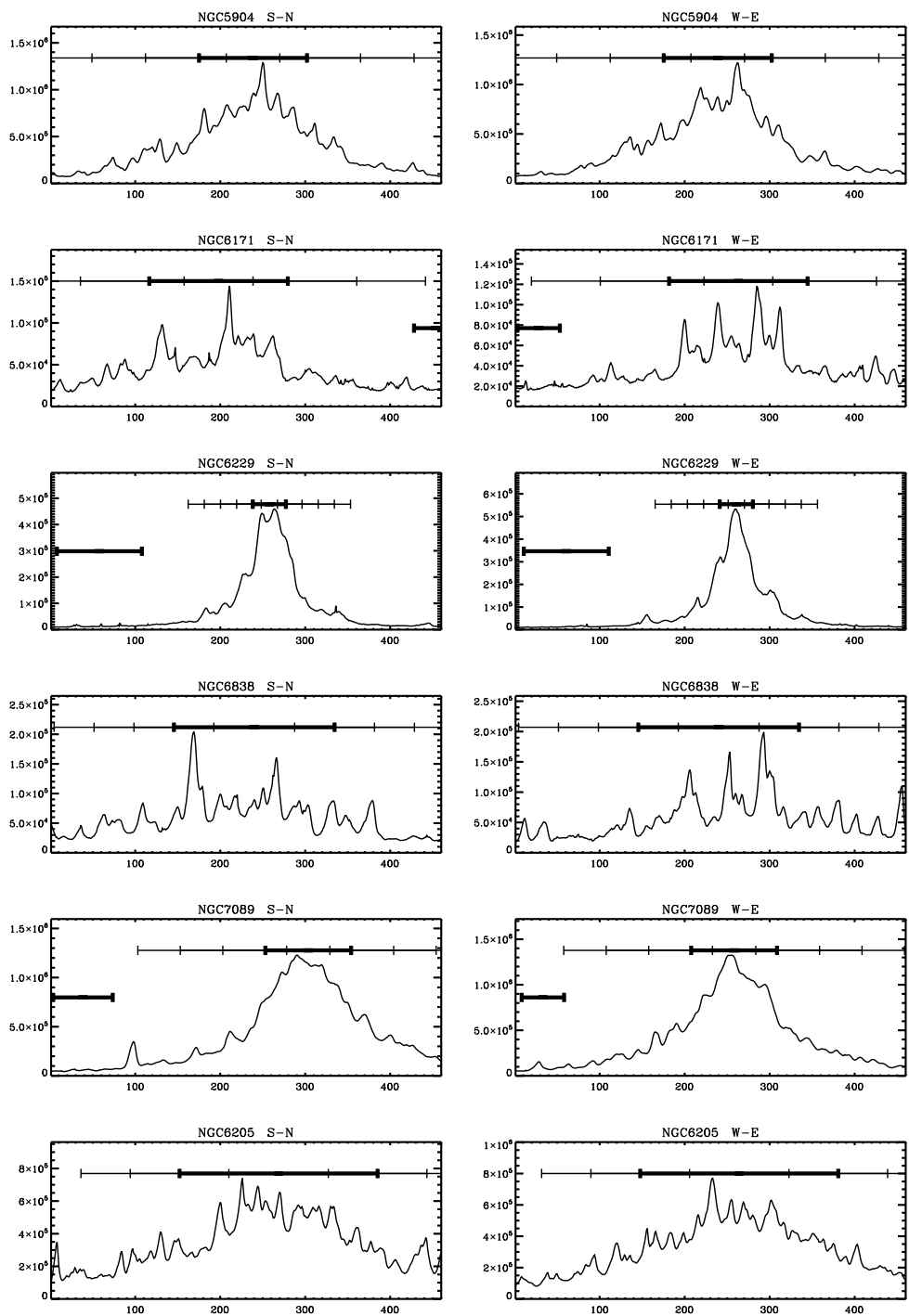}
\caption{Observed light profiles of the 24 GGCs. The profiles were obtained by adding flux along the  dispersion axis. The x--axis is in pixels and the y--axis is in flux units. In each panel, the upper bars indicate various extraction windows with the thick upper bar equal to the size of the core radius. Lower thick bars represent extraction windows of the background sky selected well away from the globular cluster.}
  \label{fig:gc_apr}
 \end{center} 
\end{figure}

\clearpage
\addtocounter{figure}{-1}
\begin{figure}
\begin{center}
\includegraphics[scale=0.95]{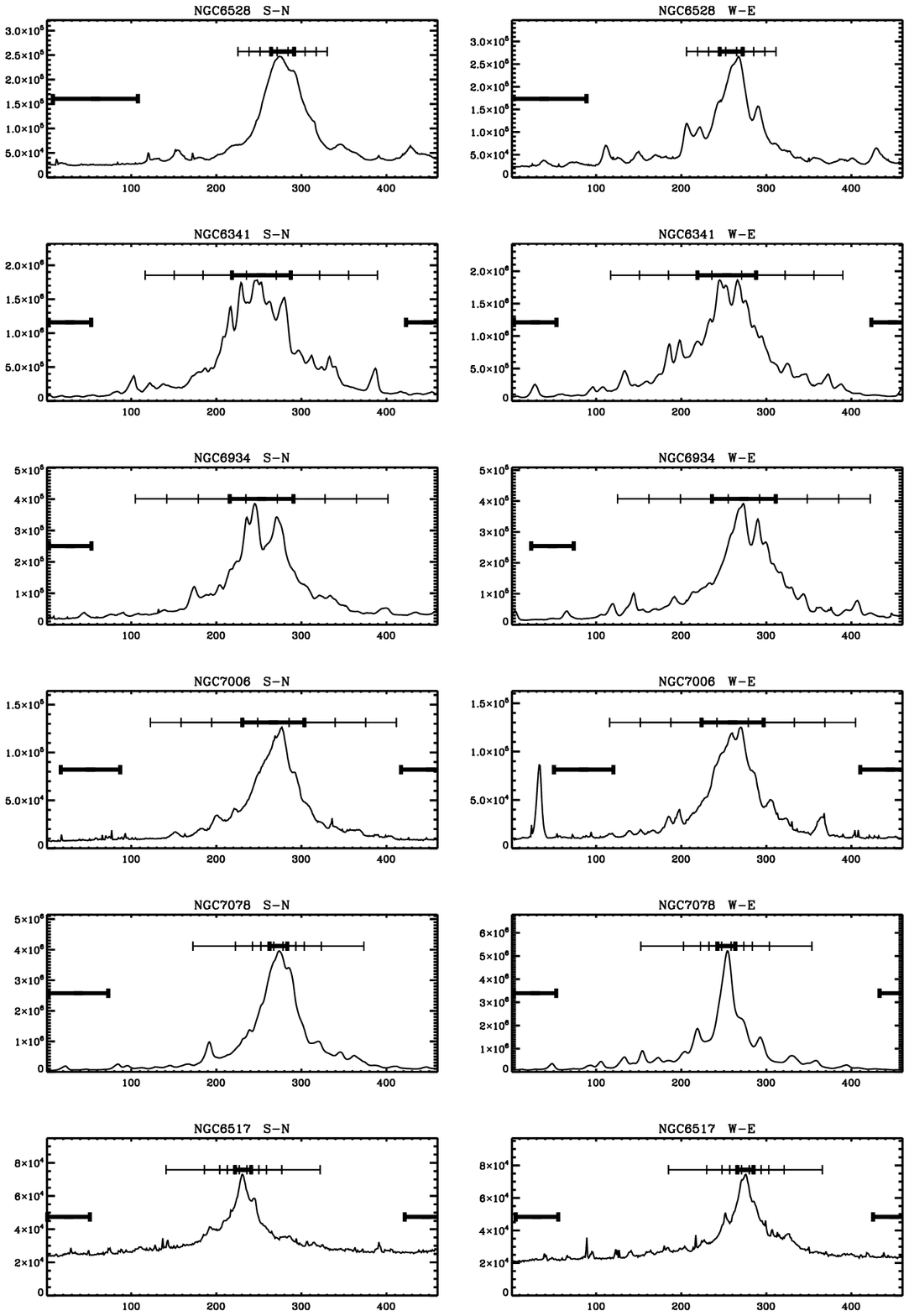}
\caption[Continued]{Continued.}
\end{center} 
\end{figure}

\clearpage
\addtocounter{figure}{-1}
\begin{figure}
\begin{center}
\includegraphics[scale=0.95]{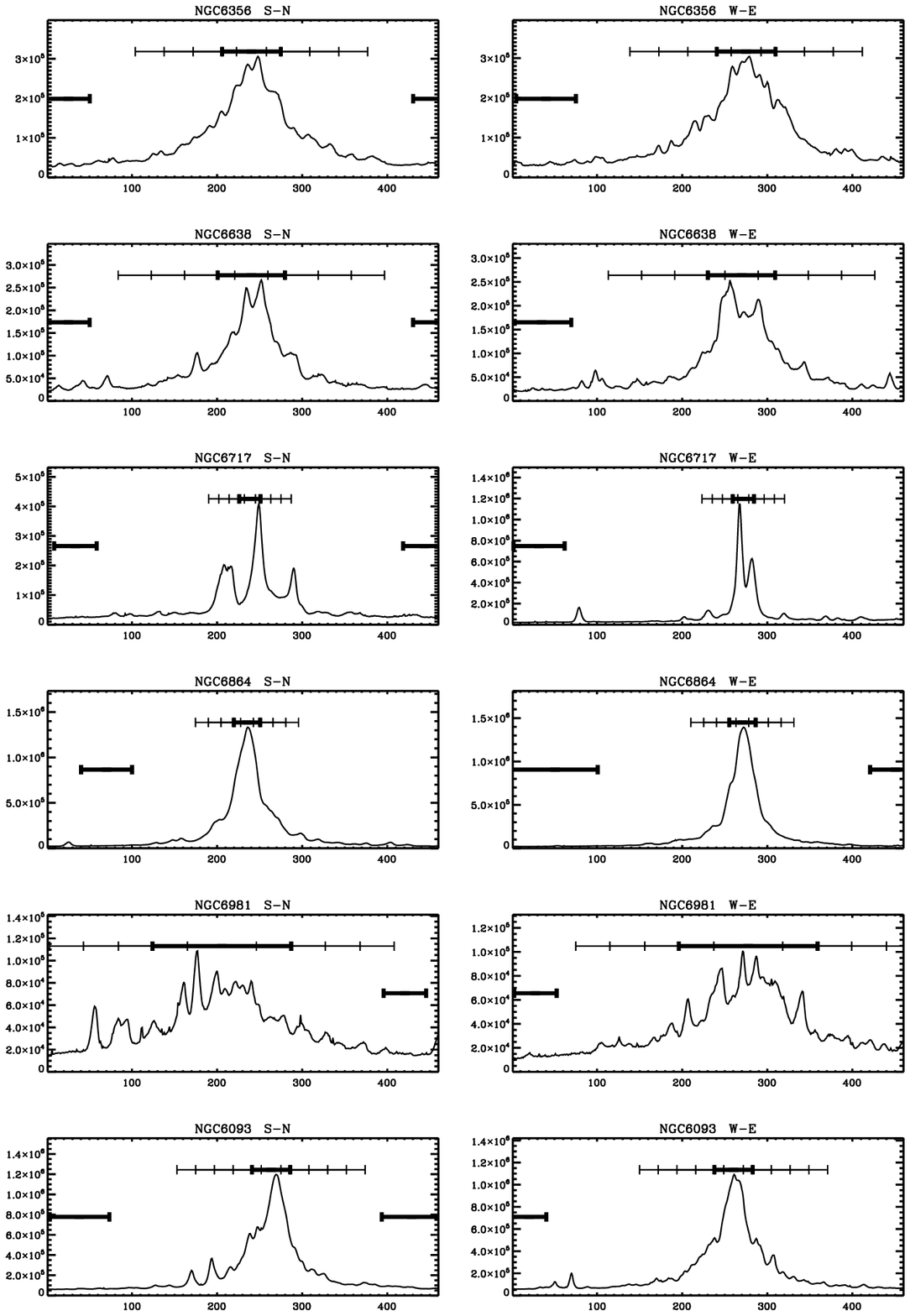}
\caption[Continued]{Continued.}
 \end{center} 
\end{figure}

\clearpage
\addtocounter{figure}{-1}
\begin{figure}
 \begin{center}
\includegraphics[scale=0.95]{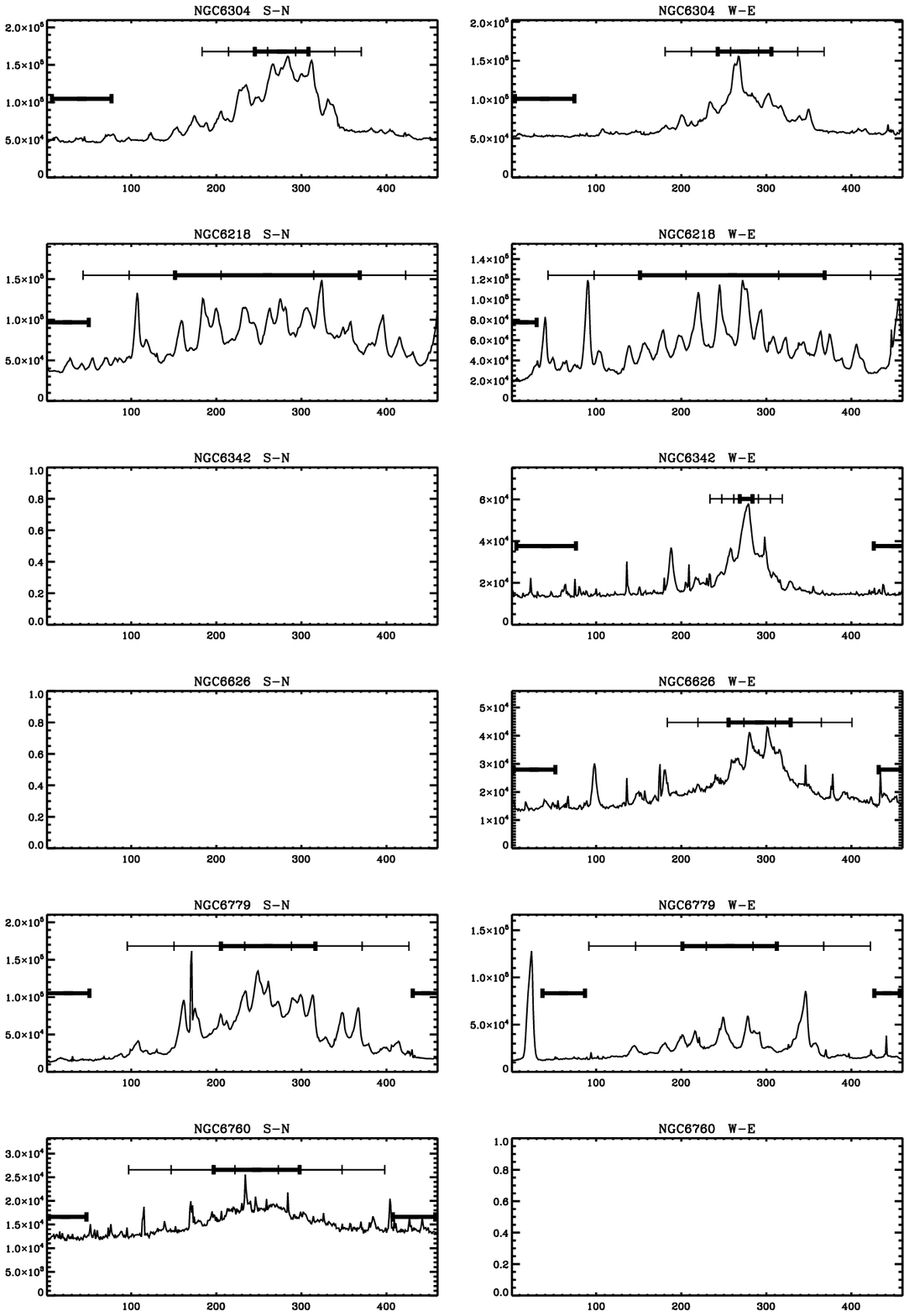}
\caption[Continued]{Continued.}
  \end{center} 
\end{figure}

\clearpage
\begin{figure}
 \begin{center}
\includegraphics[scale=0.90]{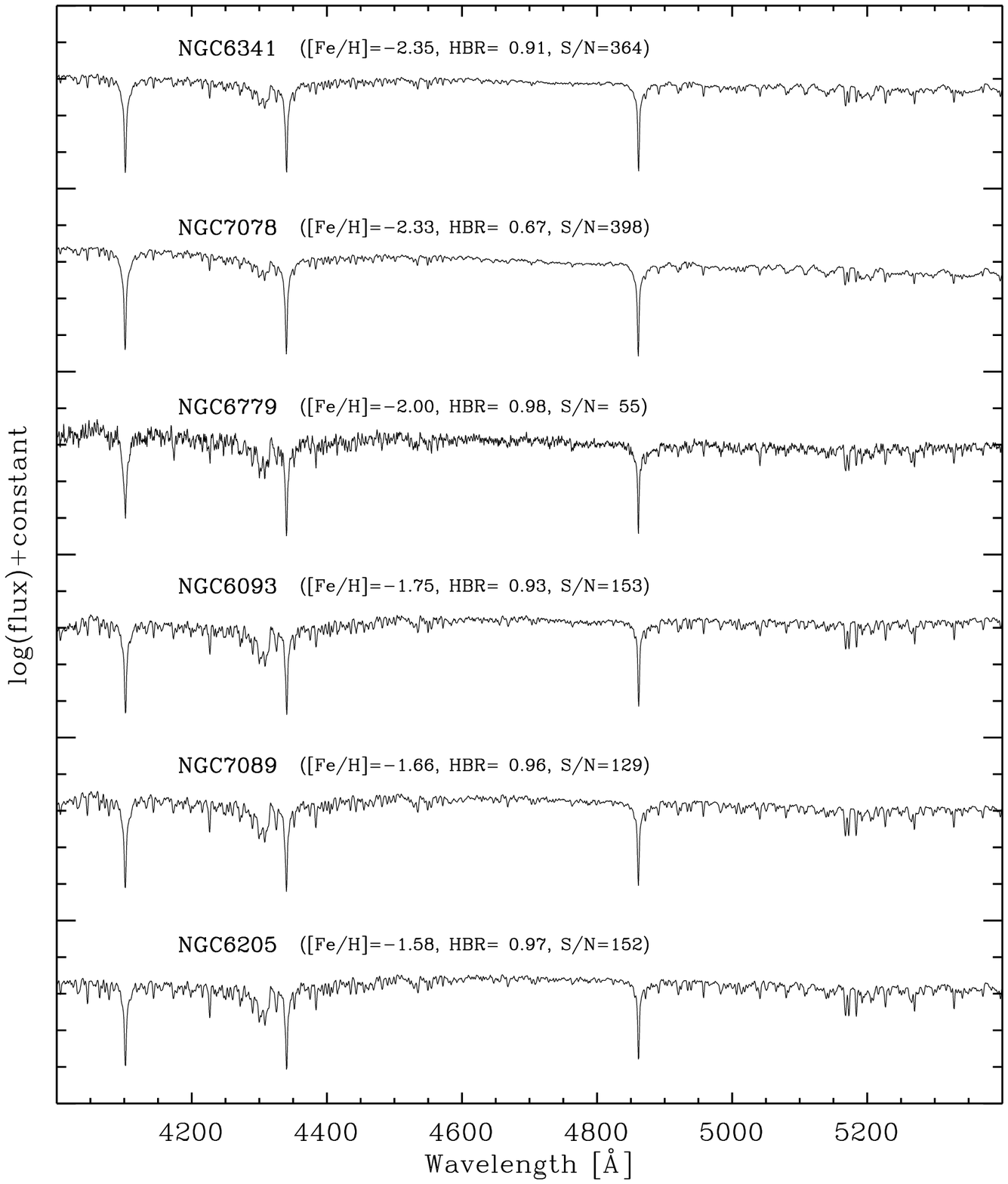}
\caption{Flux-calibrated, dereddened, combined spectra of the observed GGCs. The spectra are ordered by metallicity and shifted vertically by arbitrary constants for clarity. GC names with metallicity, horizontal-branch ratio, and signal-to-noise ratio near 4700\,\AA~are denoted above the respective spectra.}
  \label{fig:spec_gcs}
 \end{center} 
\end{figure}

\clearpage
\addtocounter{figure}{-1}
\begin{figure}
 \begin{center}
\includegraphics[scale=0.90]{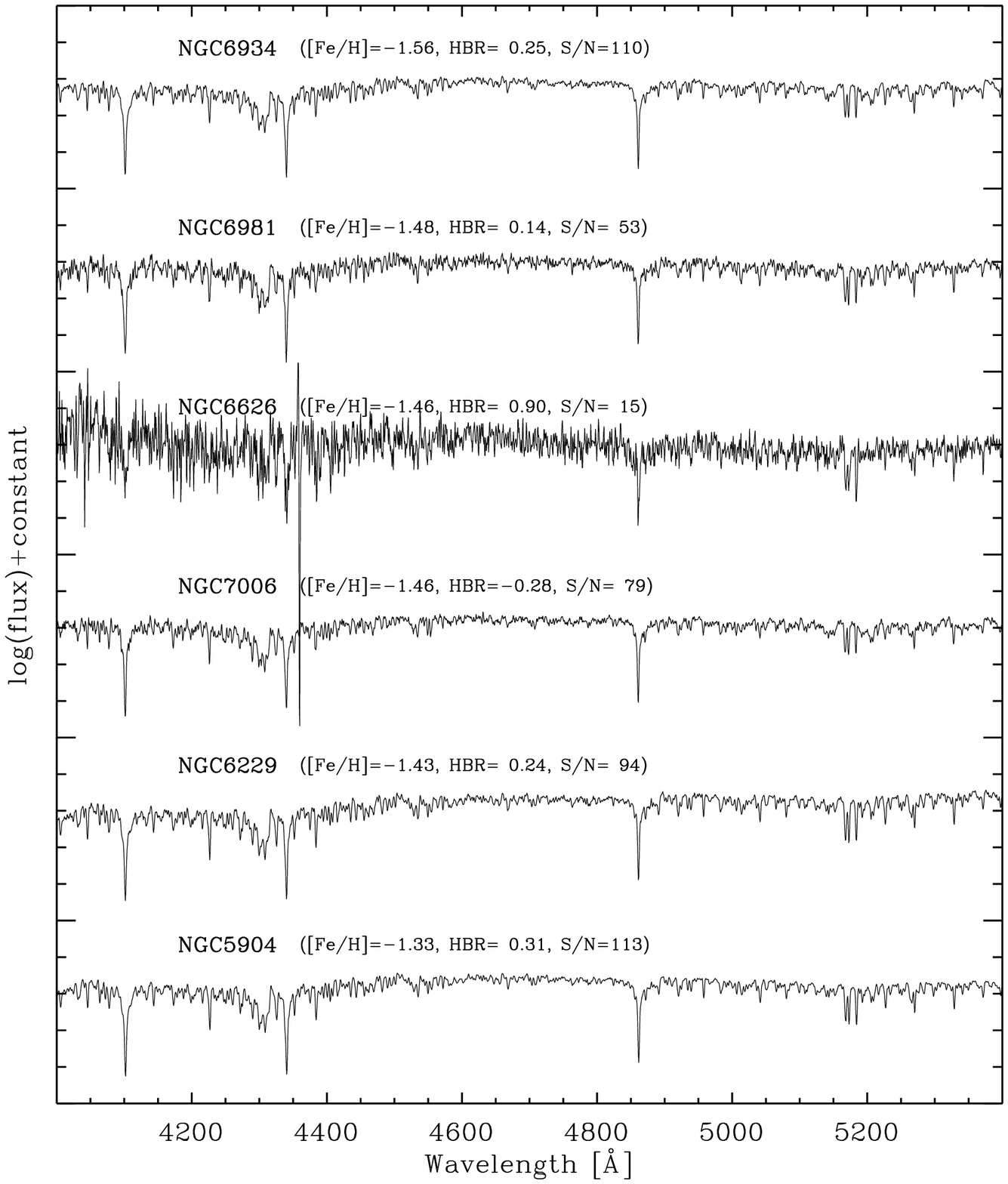}
\caption{continued.}
  \label{fig:com_pu}
 \end{center} 
\end{figure}

\clearpage
\addtocounter{figure}{-1}
\begin{figure}
 \begin{center}
\includegraphics[scale=0.90]{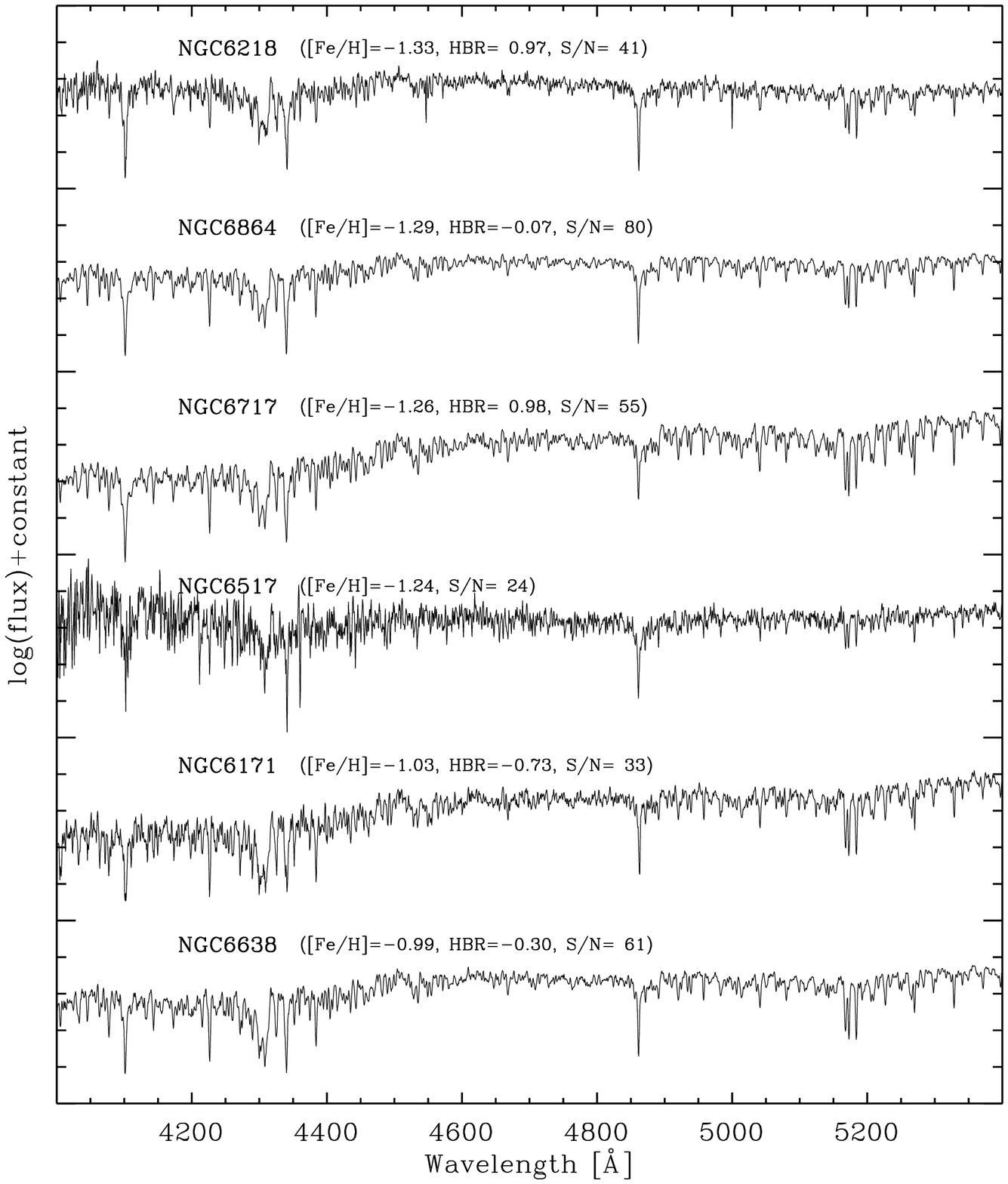}
\caption{continued.}
  \label{fig:com_pu}
 \end{center} 
\end{figure}

\clearpage
\addtocounter{figure}{-1}
\begin{figure}
 \begin{center}
\includegraphics[scale=0.90]{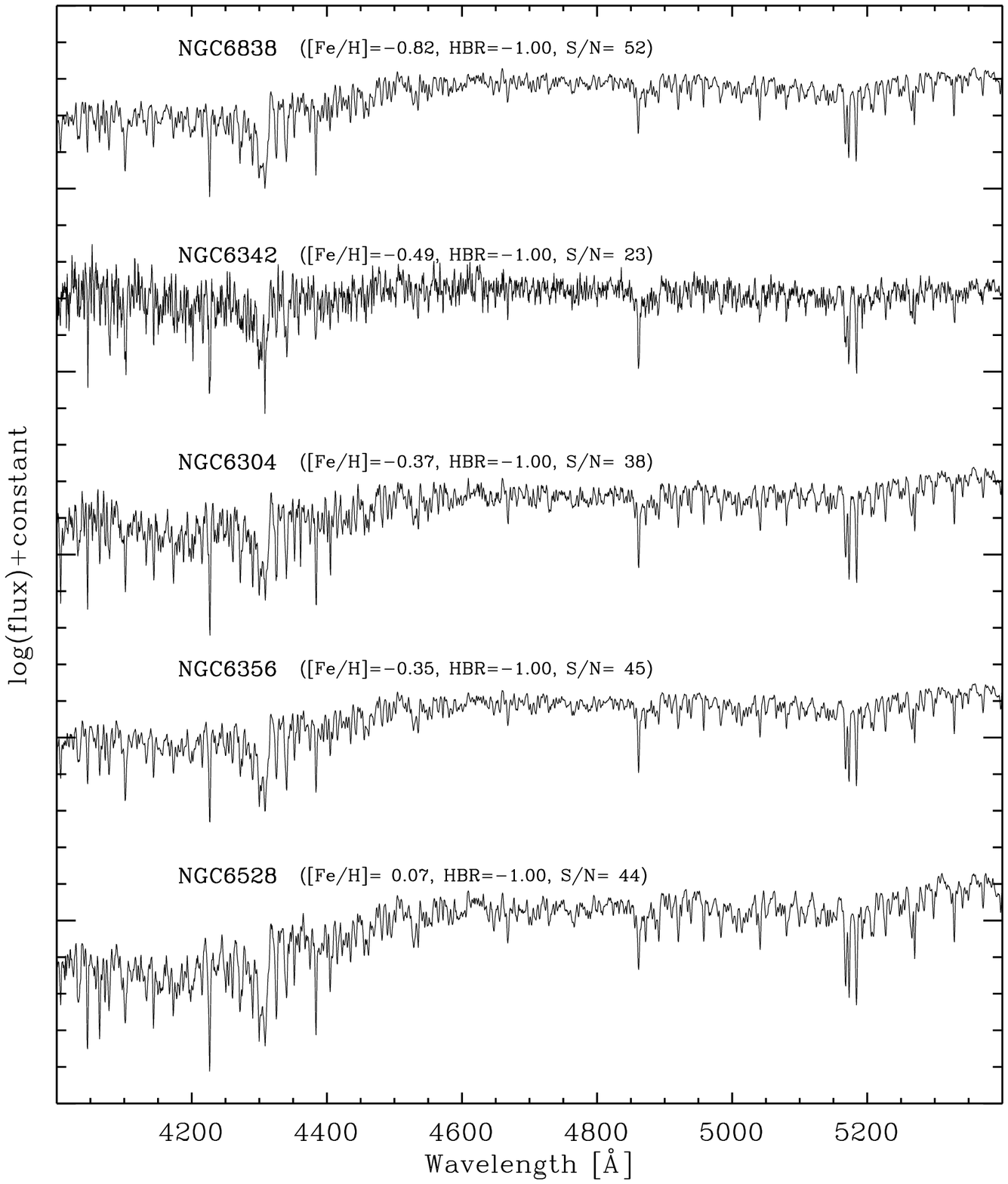}
\caption{continued.}
  \label{fig:com_pu}
 \end{center} 
\end{figure}

\begin{figure}
 \begin{center}
\includegraphics{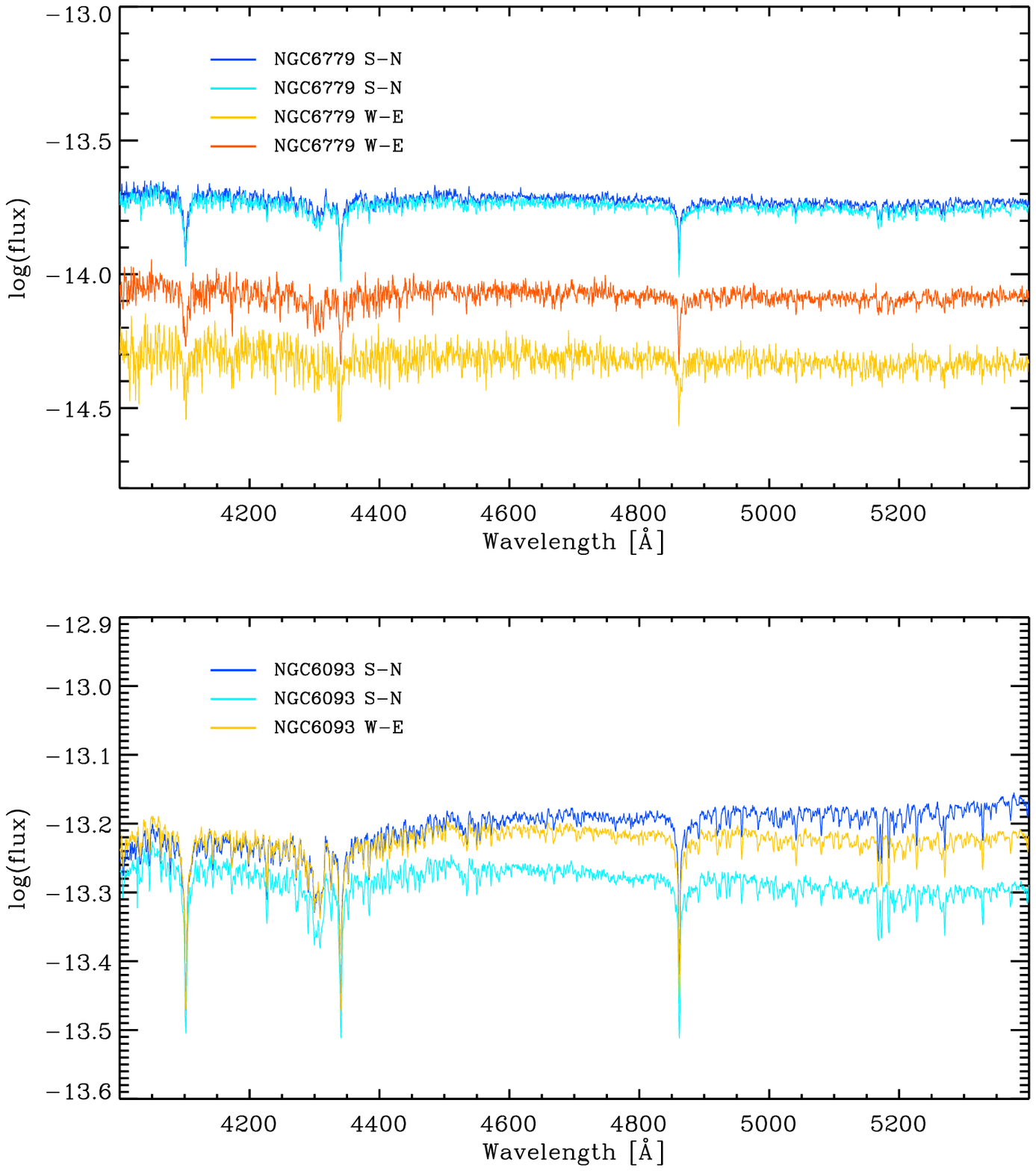}
\caption{Differences in spectral shape between individual spectra for the same GC. The spectra of NGC 6779 (upper panel) show a difference in continuum flux level between exposures, while the spectra of NGC 6093 show a difference in the spectral slope.}
  \label{fig:spec_gcs2}
 \end{center} 
\end{figure}

\clearpage
\begin{deluxetable}{crr}
\tablecolumns{3}
\tablecaption{Radial velocities of sample GCs}
\tablewidth{0pt}
\tablehead{\colhead{GC} & \colhead{$V\tablenotemark{a}_{r,this~work}$[km/s]} & \colhead{$V\tablenotemark{b}_{r,ref}$[km/s]}}
\startdata
 NGC~5904 &   53.9 $\pm$ 3.5 &   53.2 $\pm$  0.4 \\
 NGC~6093 &    4.2 $\pm$ 3.5 &    8.1 $\pm$  1.5 \\
 NGC~6171 &   --42.6 $\pm$ 4.0 &   --34.1 $\pm$  0.3 \\
 NGC~6205 &  --244.8 $\pm$ 3.7 &  --244.2 $\pm$  0.2 \\
 NGC~6218 &   --56.7 $\pm$ 3.1 &   --41.4 $\pm$  0.2 \\
 NGC~6229 &  --141.5 $\pm$ 7.0 &  --154.2 $\pm$  7.6 \\
 NGC~6304 &  --104.8 $\pm$ 3.3 &  --107.3 $\pm$  3.6 \\
 NGC~6341 &  --105.2 $\pm$ 5.8 &  --120.0 $\pm$  0.1 \\
 NGC~6342 &  128.9 $\pm$ 4.2 &  115.7 $\pm$  1.4 \\
 NGC~6356 &   42.2 $\pm$ 3.5 &   27.0 $\pm$  4.3 \\
 NGC~6517 &   --34.0 $\pm$ 4.6 &   --39.6 $\pm$  8.0 \\
 NGC~6528 &  208.7 $\pm$ 5.0 &  206.6 $\pm$  1.4 \\
 NGC~6626 &   13.8 $\pm$ 8.9 &   17.0 $\pm$  1.0 \\
 NGC~6638 &   11.2 $\pm$ 3.5 &   18.1 $\pm$  3.9 \\
 NGC~6717 &   28.7 $\pm$ 3.7 &   22.8 $\pm$  3.4 \\
 NGC~6760 &   --38.1 $\pm$ 9.9 &   --27.5 $\pm$  6.3 \\
 NGC~6779 &  --138.2 $\pm$ 3.4 &  --135.6 $\pm$  0.9 \\
 NGC~6838 &   --21.4 $\pm$ 3.5 &   --22.8 $\pm$  0.2 \\
 NGC~6864 &  --184.6 $\pm$ 3.8 &  --189.3 $\pm$  3.6 \\
 NGC~6934 &  --406.7 $\pm$ 3.9 &  --411.4 $\pm$  1.6 \\
 NGC~6981 &  --333.5 $\pm$ 4.2 &  --345.0 $\pm$  3.7 \\
 NGC~7006 &  --383.2 $\pm$ 4.3 &  --384.1 $\pm$  0.4 \\
 NGC~7078 &   --98.4 $\pm$ 3.7 &  --107.0 $\pm$  0.2 \\
 NGC~7089 &    --6.4 $\pm$ 3.5 &    --5.3 $\pm$  2.0 \\
\enddata
\vspace{-4mm}
\tablenotetext{a}{This work}
\tablenotetext{b}{Taken from \citet{har96}}
\label{tab:mwgc_rd}
\end{deluxetable}

\clearpage
\begin{figure}
\begin{center}
\includegraphics[scale=1.0]{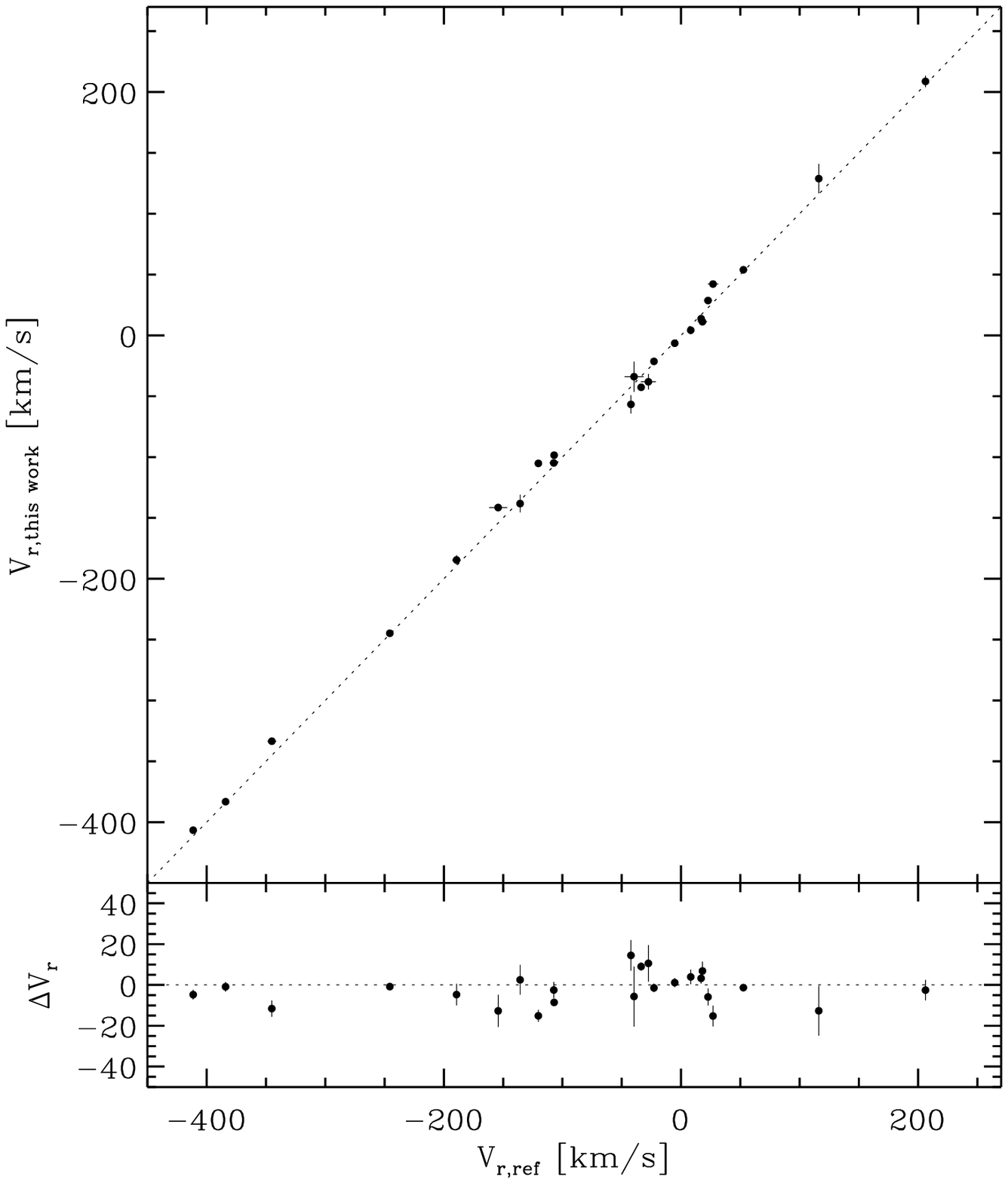}
\caption{Galactic globular cluster radial velocity comparisons. In the upper panel, our measurements are on the y--axis, and the values from \citet{har96} are on the x--axis. The dotted line represents a one-to-one relationship. The data points are listed in Table \ref{tab:mwgc_rd}. In the lower panel, the differences of the two measurements, $\Delta V_{r}=V_{r,ref}-V_{r,this\: work}$, are plotted.}
  \label{fig:com_gcvel}
 \end{center} 
\end{figure}

\begin{figure}
\begin{center}
\includegraphics[scale=0.9]{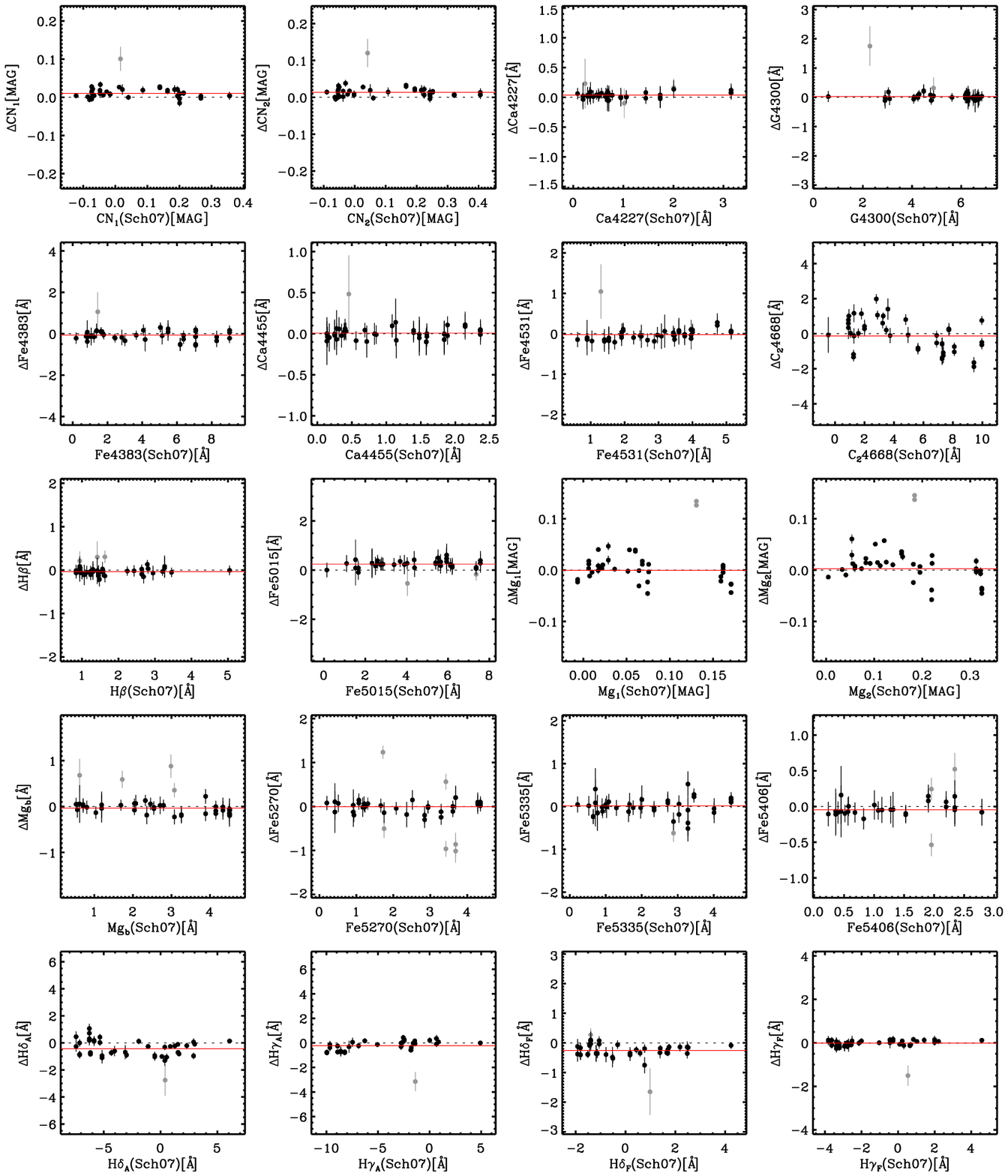}
\caption{Comparison of 20 Lick index measurements for the calibration stars with those reported in S07. The red solid lines represent the offsets between the two datasets, which are determined by calculating the error-weighted mean values of the index measurement differences after removing outliers (gray dots) with three-sigma clipping.}
  \label{fig:lick_cal}
 \end{center} 
\end{figure}


\clearpage
\begin{deluxetable}{lr}
\tablecolumns{2}
\tabletypesize{\footnotesize}
\tablecaption{Lick index offsets between ours and the S07 data}
\tablewidth{0pt}
\tablehead{\colhead{Line Index} & \colhead{$\Delta_{Lick}$\tablenotemark{a}}}
\startdata
     CN${_1}$ &  0.0105 $\pm$ 0.0009 mag \\
     CN${_2}$ &  0.0135 $\pm$ 0.0011 mag \\
       Ca4227 &    0.04 $\pm$   0.02 \AA \\
        G4300 &    0.03 $\pm$   0.03 \AA \\
       Fe4383 &   --0.07 $\pm$   0.04 \AA \\
       Ca4455 &    0.01 $\pm$   0.02 \AA \\
       Fe4531 &   --0.02 $\pm$   0.03 \AA \\
  C${_2}$4668 &   --0.13 $\pm$   0.05 \AA \\
     H$\beta$ &   --0.04 $\pm$   0.02 \AA \\
       Fe5015 &    0.24 $\pm$   0.05 \AA \\
     Mg${_1}$ & --0.0004 $\pm$ 0.0005 mag \\
     Mg${_2}$ &  0.0027 $\pm$ 0.0006 mag \\
     Mg${_b}$ &  --0.037 $\pm$  0.025 \AA \\
       Fe5270 &   --0.01 $\pm$   0.03 \AA \\
       Fe5335 &    0.01 $\pm$   0.03 \AA \\
       Fe5406 &   --0.04 $\pm$   0.03 \AA \\
H${\delta_A}$ &   --0.43 $\pm$   0.03 \AA \\
H${\gamma_A}$ &   --0.22 $\pm$   0.04 \AA \\
H${\delta_F}$ &   --0.25 $\pm$   0.02 \AA \\
H${\gamma_F}$ &   --0.01 $\pm$   0.02 \AA \\
\enddata
\tablenotetext{a}{$EW_{S07}-EW_{This~work}$}
\label{tab:lick_off}
\end{deluxetable}

\begin{figure}
\begin{center}
\includegraphics[scale=0.9]{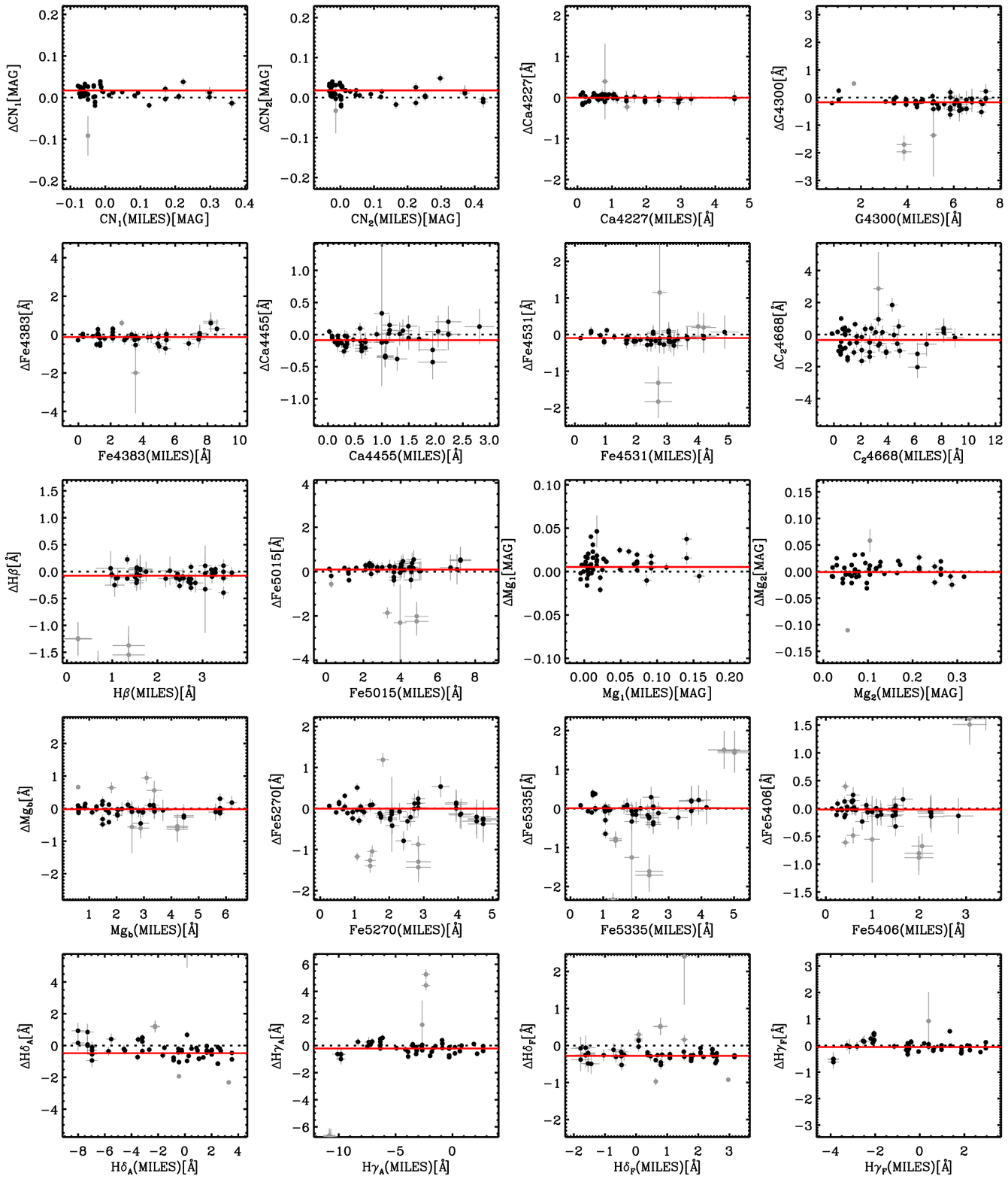}
\caption{Comparison of 20 Lick index measurements on the LIS system for the calibration stars with the measurements on the MILES published spectra. The red solid lines represent the offsets between our data and the MILES data, which are determined by calculating the error-weighted mean values of the index measurement differences after removing outliers (gray dots) with three-sigma clipping.}
  \label{fig:lis_cal}
 \end{center} 
\end{figure}

\clearpage
\begin{deluxetable}{lr}
\tablecolumns{2}
\tabletypesize{\footnotesize}
\tablecaption{LIS index offsets between ours and the MILES data}
\tablewidth{0pt}
\tablehead{\colhead{Line Index} & \colhead{$\Delta_{LIS}$\tablenotemark{a}}}
\startdata
     CN${_1}$ &  0.0170 $\pm$ 0.0004 mag \\
     CN${_2}$ &  0.0180 $\pm$ 0.0006 mag \\
       Ca4227 &   --0.00 $\pm$   0.01 \AA \\
        G4300 &   --0.17 $\pm$   0.02 \AA \\
       Fe4383 &   --0.13 $\pm$   0.03 \AA \\
       Ca4455 &   --0.09 $\pm$   0.01 \AA \\
       Fe4531 &   --0.09 $\pm$   0.02 \AA \\
  C${_2}$4668 &   --0.34 $\pm$   0.03 \AA \\
     H$\beta$ &   --0.08 $\pm$   0.01 \AA \\
       Fe5015 &    0.09 $\pm$   0.03 \AA \\
     Mg${_1}$ &  0.0053 $\pm$ 0.0003 mag \\
     Mg${_2}$ & --0.0010 $\pm$ 0.0004 mag \\
     Mg${_b}$ &  --0.018 $\pm$  0.015 \AA \\
       Fe5270 &   --0.00 $\pm$   0.02 \AA \\
       Fe5335 &    0.01 $\pm$   0.02 \AA \\
       Fe5406 &   --0.02 $\pm$   0.02 \AA \\
H${\delta_A}$ &   --0.48 $\pm$   0.02 \AA \\
H${\gamma_A}$ &   --0.22 $\pm$   0.02 \AA \\
H${\delta_F}$ &   --0.28 $\pm$   0.01 \AA \\
H${\gamma_F}$ &   --0.05 $\pm$   0.01 \AA \\
\enddata
\tablenotetext{a}{$EW_{MILES}-EW_{This~work}$}
\label{tab:lis_off}
\end{deluxetable}

\clearpage
\begin{figure}
\begin{center}
\includegraphics[scale=0.9]{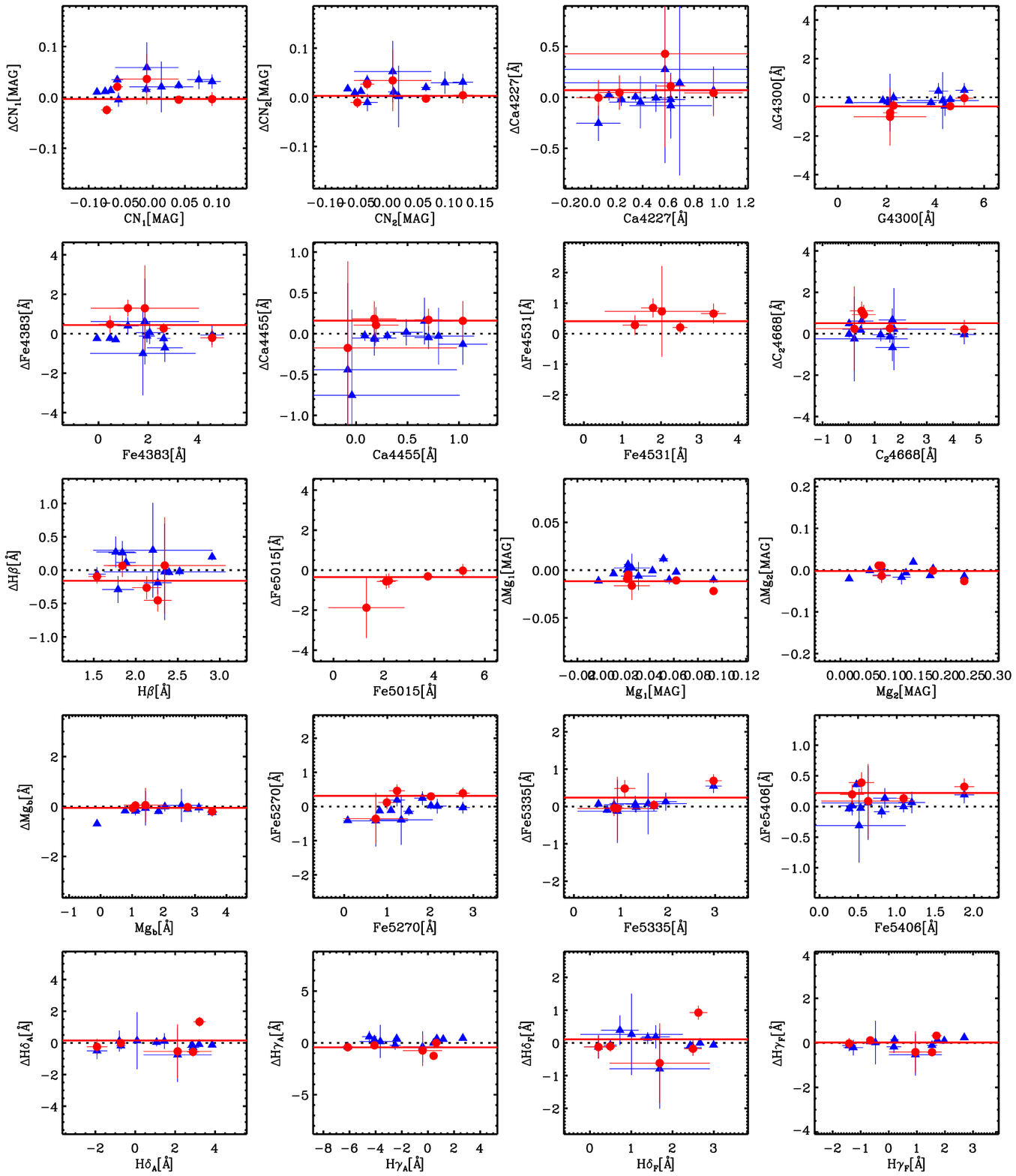}
\caption{Comparison of Lick index measurements for GGCs with previous studies. The index measurements of this study are plotted along the horizontal axis and the differences between our index values and those from PSK02 (red filled circle) and from S12 (blue filled triangle) are plotted along the vertical axis. The zero-point offset between this study and PSK02 (red solid line) is determined by the error-weighted mean value of the difference.}
  \label{fig:com_pu}
 \end{center} 
\end{figure}

\clearpage
\begin{deluxetable}{lrr}
\tablecolumns{3}
\tabletypesize{\footnotesize}
\tablecaption{Lick index offsets against previous studies}
\tablewidth{0pt}
\tablehead{\colhead{Line Index} & \colhead{$\Delta_{PSK02}$\tablenotemark{a}} & \colhead{$\Delta_{S12}$\tablenotemark{b}}}
\startdata
     CN${_1}$ & --0.0032 $\pm$ 0.0042 mag &  0.0135 $\pm$ 0.0013 mag \\
     CN${_2}$ &  0.0029 $\pm$ 0.0052 mag &  0.0142 $\pm$ 0.0018 mag \\
       Ca4227 &    0.07 $\pm$   0.08 \AA &    0.00 $\pm$   0.03 \AA \\
        G4300 &   --0.47 $\pm$   0.13 \AA &   --0.19 $\pm$   0.04 \AA \\
       Fe4383 &    0.44 $\pm$   0.18 \AA &   --0.25 $\pm$   0.07 \AA \\
       Ca4455 &    0.16 $\pm$   0.09 \AA &   --0.03 $\pm$   0.03 \AA \\
       Fe4531 &  0.4100 $\pm$ 0.1300 \AA &                      NaN \\
  C${_2}$4668 &    0.51 $\pm$   0.19 \AA &    0.19 $\pm$   0.07 \AA \\
     H$\beta$ &   --0.16 $\pm$   0.07 \AA &    0.04 $\pm$   0.03 \AA \\
       Fe5015 & --0.3500 $\pm$ 0.1500 \AA &                      NaN \\
     Mg${_1}$ & --0.0116 $\pm$ 0.0015 mag & --0.0040 $\pm$ 0.0006 mag \\
     Mg${_2}$ & --0.0021 $\pm$ 0.0018 mag & --0.0053 $\pm$ 0.0007 mag \\
     Mg${_b}$ &  --0.053 $\pm$  0.070 \AA &  --0.142 $\pm$  0.032 \AA \\
       Fe5270 &    0.31 $\pm$   0.08 \AA &   --0.16 $\pm$   0.03 \AA \\
       Fe5335 &    0.24 $\pm$   0.09 \AA &   --0.03 $\pm$   0.03 \AA \\
       Fe5406 &    0.22 $\pm$   0.06 \AA &   --0.02 $\pm$   0.03 \AA \\
H${\delta_A}$ &    0.15 $\pm$   0.15 \AA &   --0.12 $\pm$   0.05 \AA \\
H${\gamma_A}$ &   --0.43 $\pm$   0.14 \AA &    0.39 $\pm$   0.05 \AA \\
H${\delta_F}$ &    0.11 $\pm$   0.10 \AA &   --0.07 $\pm$   0.03 \AA \\
H${\gamma_F}$ &    0.02 $\pm$   0.08 \AA &    0.13 $\pm$   0.03 \AA \\
\enddata
\tablenotetext{a}{$EW_{this~work}-EW_{PSK02}$}
\tablenotetext{b}{$EW_{this~work}-EW_{S12}$}
\label{tab:gc_off}
\end{deluxetable}

\clearpage
\begin{deluxetable}{lrrrrrrrrrrr}
\rotate
\tablecolumns{12}
\tablecaption{Metallicities, Lick/S07 indices, and their uncertainties for GGCs \label{tab:Lick}}
\tablewidth{0pt}
\tablehead{\colhead{ID} & \colhead{[Fe/H]} & \colhead{CN$_{1}$} & \colhead{CN$_{2}$} & \colhead{Ca4227} & \colhead{G4300} & \colhead{Fe4383} & \colhead{Ca4455} & \colhead{Fe4531} & \colhead{C$_{2}$4668} & \colhead{H${\beta}$} & \colhead{Fe5015} \\
        & (dex) & (mag) & (mag) & ($\rm\AA$) & ($\rm\AA$) & ($\rm\AA$) & ($\rm\AA$) & ($\rm\AA$) & ($\rm\AA$) & ($\rm\AA$) & ($\rm\AA$)}
\startdata
\vspace{-2mm}
  NGC104 & --0.76 &  0.0250 &  0.0500 &  0.56 &  4.66 &  2.49 &  0.67 &   \nodata &  1.68 &  1.60 &   \nodata\\
         &  0.02 &  0.0010 &  0.0010 &  0.01 &  0.02 &  0.03 &  0.02 &   \nodata &  0.03 &  0.01 &   \nodata\\
\vspace{-2mm}
 NGC1851 & --1.18 & --0.0560 & --0.0310 &  0.34 &  2.79 &  1.28 &  0.43 &   \nodata &  0.57 &  2.27 &   \nodata\\
         &  0.08 &  0.0010 &  0.0010 &  0.02 &  0.04 &  0.05 &  0.03 &   \nodata &  0.06 &  0.02 &   \nodata\\
\vspace{-2mm}
 NGC1904 & --1.58 & --0.0850 & --0.0620 &  0.21 &  1.89 &  0.70 &  0.23 &   \nodata & --0.07 &  2.48 &   \nodata\\
         &  0.02 &  0.0020 &  0.0020 &  0.04 &  0.06 &  0.09 &  0.05 &   \nodata &  0.11 &  0.04 &   \nodata\\
\vspace{-2mm}
 NGC2298 & --1.96 & --0.1020 & --0.0790 &  0.24 &  1.66 &  0.65 &  0.15 &   \nodata &  0.17 &  2.68 &   \nodata\\
         &  0.04 &  0.0030 &  0.0040 &  0.06 &  0.10 &  0.14 &  0.07 &   \nodata &  0.16 &  0.06 &   \nodata\\
\enddata
\end{deluxetable}

\clearpage
\begin{deluxetable}{lrrrrrrrrrrc}
\rotate
\tablecolumns{12}
\addtocounter{table}{-1}
\tablecaption{Metallicities, Lick/S07 indices, and their uncertainties for GGCs, continued. \label{tab:Lick}}
\tablewidth{0pt}
\tablehead{\colhead{ID} & \colhead{Mg$_{1}$} & \colhead{Mg$_{2}$} & \colhead{Mgb} & \colhead{Fe5270} & \colhead{Fe5335} & \colhead{Fe5406} & \colhead{H${\delta_A}$} & \colhead{H${\gamma_A}$} & \colhead{H${\delta_F}$} & \colhead{H${\gamma_F}$} & \colhead{Source$^{a}$} \\
     & (mag) & (mag) & ($\rm\AA$) & ($\rm\AA$) & ($\rm\AA$) & ($\rm\AA$) & ($\rm\AA$) & ($\rm\AA$) & ($\rm\AA$) & ($\rm\AA$)}
\startdata
\vspace{-2mm}
  NGC104 &  0.0569 &  0.1589 &  2.690 &  1.91 &  1.68 &  1.06 & --0.48 & --4.27 &  0.68 & --0.66 &    2\\
         &  0.0003 &  0.0004 &  0.010 &  0.02 &  0.02 &  0.01 &  0.02 &  0.02 &  0.02 &  0.01   \\
\vspace{-2mm}
 NGC1851 &  0.0251 &  0.0829 &  1.350 &  1.40 &  1.12 &  0.67 &  2.23 & --0.60 &  2.06 &  1.23 &    2\\
         &  0.0005 &  0.0006 &  0.020 &  0.03 &  0.03 &  0.02 &  0.04 &  0.03 &  0.02 &  0.02   \\
\vspace{-2mm}
 NGC1904 &  0.0103 &  0.0485 &  0.830 &  0.93 &  0.78 &  0.34 &  3.17 &  0.87 &  2.61 &  1.88 &    2\\
         &  0.0009 &  0.0011 &  0.050 &  0.05 &  0.06 &  0.04 &  0.06 &  0.06 &  0.04 &  0.04   \\
\vspace{-2mm}
 NGC2298 &  0.0070 &  0.0442 &  0.790 &  0.74 &  0.63 &  0.24 &  3.70 &  1.45 &  2.87 &  2.13 &    2\\
         &  0.0013 &  0.0016 &  0.070 &  0.07 &  0.08 &  0.06 &  0.09 &  0.09 &  0.07 &  0.06   \\
\enddata
\vspace{-8mm}
\tablecomments{(a) Sources of Lick indices: 1 = this work; 2 = S12; 3 = PSK02 (calibrated to this work)}
\end{deluxetable}

\clearpage
\begin{deluxetable}{lrrrrrrrrrrr}
\rotate
\tablecolumns{12}
\tablecaption{Metallicities, LIS indices, and their uncertainties for GGCs \label{tab:LIS}}
\tablewidth{0pt}
\tablehead{\colhead{ID} & \colhead{[Fe/H]} & \colhead{CN$_{1}$} & \colhead{CN$_{2}$} & \colhead{Ca4227} & \colhead{G4300} & \colhead{Fe4383} & \colhead{Ca4455} & \colhead{Fe4531} & \colhead{C$_{2}$4668} & \colhead{H${\beta}$} & \colhead{Fe5015} \\
        & (dex) & (mag) & (mag) & ($\rm\AA$) & ($\rm\AA$) & ($\rm\AA$) & ($\rm\AA$) & ($\rm\AA$) & ($\rm\AA$) & ($\rm\AA$) & ($\rm\AA$)}
\startdata
\vspace{-2mm}
  NGC104 & --0.76 &  0.0520 &  0.0931 &  0.82 &  4.78 &  2.76 &  0.77 &   \nodata &  1.75 &  1.56 &   \nodata\\
         &  0.02 &  0.0010 &  0.0020 &  0.03 &  0.05 &  0.07 &  0.04 &   \nodata &  0.08 &  0.03 &   \nodata\\
\vspace{-2mm}
 NGC1851 & --1.18 & --0.0320 &  0.0111 &  0.46 &  2.92 &  1.27 &  0.47 &   \nodata &  0.56 &  2.27 &   \nodata\\
         &  0.08 &  0.0020 &  0.0030 &  0.04 &  0.08 &  0.11 &  0.06 &   \nodata &  0.13 &  0.05 &   \nodata\\
\vspace{-2mm}
 NGC1904 & --1.58 & --0.0620 & --0.0209 &  0.26 &  1.97 &  0.59 &  0.21 &   \nodata & --0.15 &  2.49 &   \nodata\\
         &  0.02 &  0.0017 &  0.0024 &  0.04 &  0.06 &  0.09 &  0.05 &   \nodata &  0.11 &  0.04 &   \nodata\\
\vspace{-2mm}
 NGC2298 & --1.96 & --0.0809 & --0.0379 &  0.27 &  1.68 &  0.47 &  0.14 &   \nodata &  0.10 &  2.69 &   \nodata\\
         &  0.04 &  0.0048 &  0.0057 &  0.09 &  0.17 &  0.24 &  0.12 &   \nodata &  0.28 &  0.10 &   \nodata\\
\enddata
\end{deluxetable}

\clearpage
\begin{deluxetable}{lrrrrrrrrrrc}
\rotate
\tablecolumns{12}
\addtocounter{table}{-1}
\tablecaption{Metallicities, LIS indices, and their uncertainties for GGCs, continued. \label{tab:LIS}}
\tablewidth{0pt}
\tablehead{\colhead{ID} & \colhead{Mg$_{1}$} & \colhead{Mg$_{2}$} & \colhead{Mgb} & \colhead{Fe5270} & \colhead{Fe5335} & \colhead{Fe5406} & \colhead{H${\delta_A}$} & \colhead{H${\gamma_A}$} & \colhead{H${\delta_F}$} & \colhead{H${\gamma_F}$} & \colhead{Source$^{a}$} \\
     & (mag) & (mag) & ($\rm\AA$) & ($\rm\AA$) & ($\rm\AA$) & ($\rm\AA$) & ($\rm\AA$) & ($\rm\AA$) & ($\rm\AA$) & ($\rm\AA$)}
\startdata
\vspace{-2mm}
  NGC104 &  0.0596 &  0.1515 &  2.540 &  1.99 &  1.96 &  1.29 & --0.93 & --3.86 &  0.69 & --0.62 &    2\\
         &  0.0010 &  0.0010 &  0.030 &  0.03 &  0.04 &  0.03 &  0.05 &  0.05 &  0.04 &  0.03   \\
\vspace{-2mm}
 NGC1851 &  0.0276 &  0.0755 &  1.140 &  1.44 &  1.34 &  0.85 &  1.94 & --0.27 &  2.11 &  1.35 &    2\\
         &  0.0010 &  0.0010 &  0.054 &  0.06 &  0.07 &  0.05 &  0.08 &  0.08 &  0.05 &  0.05   \\
\vspace{-2mm}
 NGC1904 &  0.0118 &  0.0405 &  0.622 &  0.92 &  0.92 &  0.49 &  2.96 &  1.23 &  2.73 &  2.03 &    2\\
         &  0.0009 &  0.0009 &  0.044 &  0.05 &  0.06 &  0.04 &  0.06 &  0.06 &  0.04 &  0.04   \\
\vspace{-2mm}
 NGC2298 &  0.0093 &  0.0364 &  0.570 &  0.72 &  0.74 &  0.36 &  3.53 &  1.81 &  3.00 &  2.29 &    2\\
         &  0.0024 &  0.0024 &  0.100 &  0.12 &  0.13 &  0.10 &  0.16 &  0.16 &  0.11 &  0.10   \\
\enddata
\vspace{-8mm}
\tablecomments{(a) Sources of LIS indices: 1 = this work; 2 = SRC05}
\end{deluxetable}

\clearpage
\begin{figure}
 \begin{center}
\includegraphics[scale=0.85]{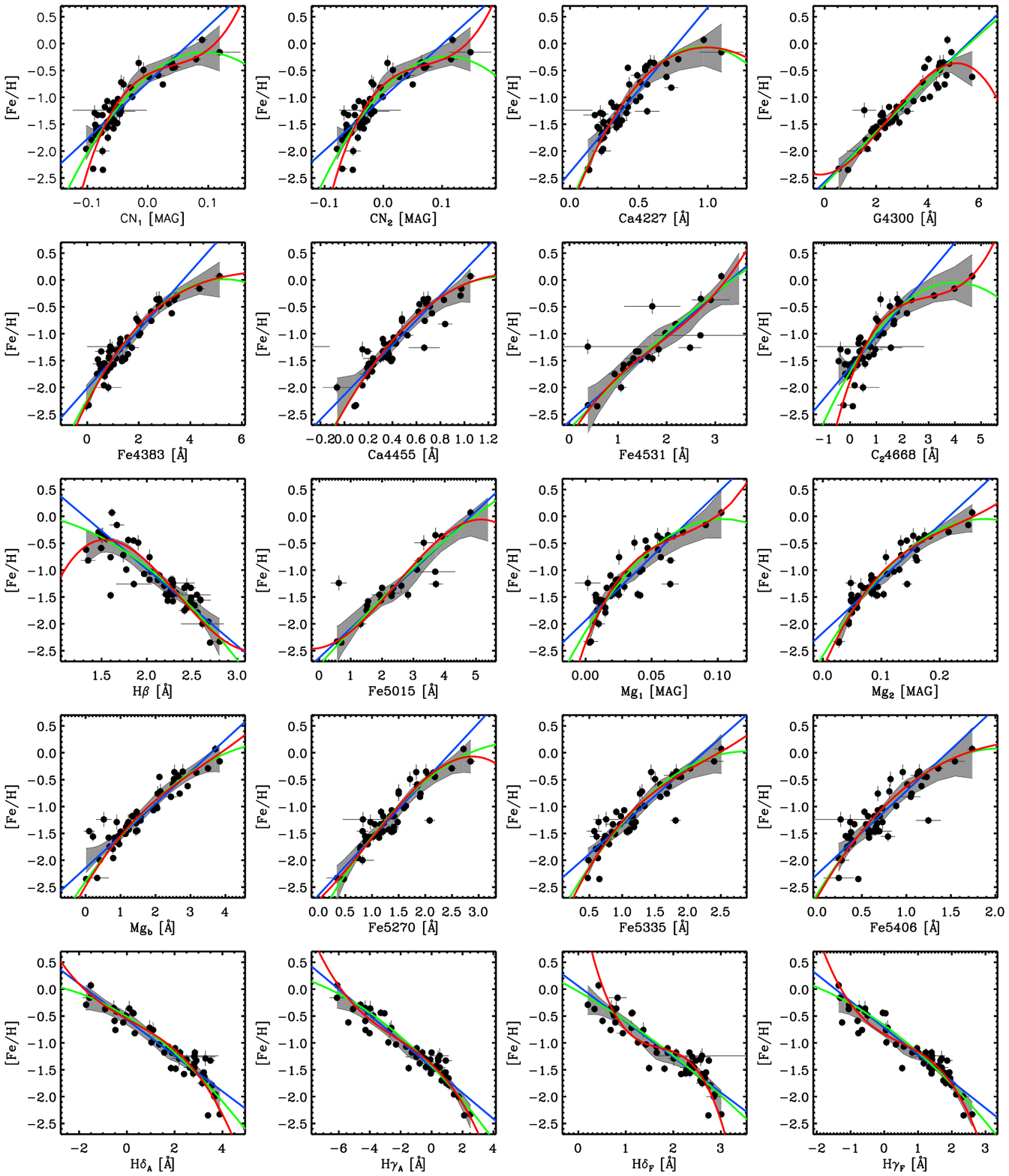}
\caption{The Lick/S07 index--metallicity relations. 
The blue, green, and red solid lines represent the first-, second-, and third-order orthogonal least-squares fit, respectively. The gray-shaded regions indicate the 95\% confidence bands based on the LOESS regression, which show the underlying trend of the data.}
  \label{fig:Indmet}
 \end{center} 
\end{figure}

\clearpage
\begin{deluxetable}{lrrrr}
\tablecolumns{5}
\tabletypesize{\small}
\rotate
\tablecaption{Coefficients of the Lick/S07 index versus metallicity relations. The coefficients were obtained by the orthogonal distance least-squares fit of a linear function $\rm[Fe/H]\it=a_{0}+a_{1} \times (index)$ to 53 GGC data.}
\tablewidth{0pt}
\tablehead{\colhead{Index} & \colhead{$a_{0}$} & \colhead{$a_{1}$} & \colhead{BIC\tablenotemark{a}} & \colhead{Valid Range}}
\startdata
     CN$_{1}$~[mag] &~~~  --0.7236 $\pm$  0.0414 ~~~&~~~ 10.4914 $\pm$  0.6798 ~~~& 59.5 & [  --1.4887,  2.4745 ] \\
     CN$_{2}$~[mag] &~~~  --1.0003 $\pm$  0.0319 ~~~&~~~  9.8522 $\pm$  0.6217 ~~~& 53.1 & [  --2.0814,  3.0346 ] \\
       Ca4227~[\AA] &~~~  --2.3964 $\pm$  0.0908 ~~~&~~~  3.0696 $\pm$  0.2045 ~~~& 71.4 & [  --2.08,  3.03 ] \\
        G4300~[\AA] &~~~  --2.5737 $\pm$  0.0670 ~~~&~~~  0.4640 $\pm$  0.0208 ~~~& 27.7 & [  --1.49,  2.45 ] \\
       Fe4383~[\AA] &~~~  --2.0139 $\pm$  0.0508 ~~~&~~~  0.5412 $\pm$  0.0274 ~~~& 31.5 & [  --2.08,  3.02 ] \\
       Ca4455~[\AA] &~~~  --2.1353 $\pm$  0.0546 ~~~&~~~  2.3100 $\pm$  0.1115 ~~~& 44.1 & [  --2.08,  3.02 ] \\
       Fe4531~[\AA] &~~~  --2.6327 $\pm$  0.0872 ~~~&~~~  0.7934 $\pm$  0.0417 ~~~& 14.4 & [  --2.12,  2.75 ] \\
  C$_{2}$4668~[\AA] &~~~  --1.6293 $\pm$  0.0532 ~~~&~~~  0.5806 $\pm$  0.0439 ~~~& 66.9 & [  --2.12,  3.38 ] \\
\enddata
\tablenotetext{a}{BIC=Bayesian information criterion}
\label{tab:coeff1}
\end{deluxetable}

\clearpage
\begin{deluxetable}{lrrrrr}
\tablecolumns{6}
\tabletypesize{\small}
\rotate
\tablecaption{The same as Table 9, but using a quadratic polynomial $\rm[Fe/H]\it=a_{0}+a_{1} \times (index) + a_{2} \times (index)^2$ to the 53 GGC data.}
\tablewidth{0pt}
\tablehead{\colhead{Index} & \colhead{$a_{0}$} & \colhead{$a_{1}$} & \colhead{$a_{2}$} & \colhead{BIC\tablenotemark{a}} & \colhead{Valid Range}}
\startdata
     CN$_{1}$~[mag] &  --0.6024 $\pm$  0.0581 &  9.3732 $\pm$  1.2557 &   --49.7232 $\pm$ 10.6338 & 57.1 & [  --1.4887,  2.4745 ] \\
     CN$_{2}$~[mag] &  --0.8565 $\pm$  0.0431 & 11.3391 $\pm$  0.9451 &   --52.1753 $\pm$  9.9137 & 53.4 & [  --2.0814,  3.0346 ] \\
       Ca4227~[\AA] &  --2.9955 $\pm$  0.1258 &  6.0801 $\pm$  0.4681 &    --3.1461 $\pm$  0.5158 & 56.1 & [  --2.08,  3.03 ] \\
        G4300~[\AA] &  --2.6238 $\pm$  0.1798 &  0.5025 $\pm$  0.1115 &    --0.0064 $\pm$  0.0182 & 30.9 & [  --1.49,  2.45 ] \\
       Fe4383~[\AA] &  --2.2307 $\pm$  0.0616 &  0.8459 $\pm$  0.0512 &    --0.0797 $\pm$  0.0137 & 24.7 & [  --2.08,  3.02 ] \\
       Ca4455~[\AA] &  --2.3949 $\pm$  0.0909 &  3.7176 $\pm$  0.3328 &    --1.4066 $\pm$  0.3642 & 44.3 & [  --2.08,  3.02 ] \\
       Fe4531~[\AA] &  --2.7763 $\pm$  0.1760 &  0.9818 $\pm$  0.1679 &    --0.0453 $\pm$  0.0427 &  5.4 & [  --2.12,  2.75 ] \\
  C$_{2}$4668~[\AA] &  --1.7058 $\pm$  0.0550 &  0.8308 $\pm$  0.0549 &    --0.1038 $\pm$  0.0187 & 52.1 & [  --2.12,  3.38 ] \\
\enddata
\tablenotetext{a}{BIC=Bayesian information criterion}
\label{tab:coeff2}
\end{deluxetable}

\clearpage
\begin{deluxetable}{lrrrrrr}
\tablecolumns{7}
\tabletypesize{\small}
\rotate
\tablecaption{The same as Table 9, but using a cubic polynomial $\rm[Fe/H]\it=a_{0}+a_{1} \times (index) + a_{2} \times (index)^2 + a_{3} \times (index)^3$ to the 53 GGC data.}
\tablewidth{0pt}
\tablehead{\colhead{Index} & \colhead{$a_{0}$} & \colhead{$a_{1}$} & \colhead{$a_{2}$} & \colhead{$a_{3}$} & \colhead{BIC\tablenotemark{a}} & \colhead{Valid Range}}
\startdata
     CN$_{1}$~[mag] &  --0.5734 $\pm$  0.0853 &  5.8757 $\pm$  2.5474 &    --67.1465 $\pm$    34.7753 &   544.9103 $\pm$   192.8483 & 78.0 & [  --1.4887,  2.4745 ] \\
     CN$_{2}$~[mag] &  --0.7675 $\pm$  0.0583 &  9.9262 $\pm$  1.3624 &   --101.2412 $\pm$    25.4773 &   487.9831 $\pm$   171.5745 & 74.9 & [  --2.0814,  3.0346 ] \\
       Ca4227~[\AA] &  --3.0963 $\pm$  0.2829 &  6.7401 $\pm$  1.5366 &     --4.3666 $\pm$     2.9654 &     0.6519 $\pm$     2.1818 & 60.7 & [  --2.08,  3.03 ] \\
        G4300~[\AA] &  --2.4314 $\pm$  0.3315 &  0.0733 $\pm$  0.2901 &     0.2073 $\pm$     0.0911 &     --0.0279 $\pm$     0.0099 & 38.6 & [  --1.49,  2.45 ] \\
       Fe4383~[\AA] &  --2.2833 $\pm$  0.1023 &  0.9469 $\pm$  0.1205 &     --0.1265 $\pm$     0.0537 &     0.0059 $\pm$     0.0095 & 29.5 & [  --2.08,  3.02 ] \\
       Ca4455~[\AA] &  --2.4062 $\pm$  0.1502 &  3.8081 $\pm$  0.7475 &     --1.5953 $\pm$     1.4999 &     0.1113 $\pm$     1.1220 & 48.8 & [  --2.08,  3.02 ] \\
       Fe4531~[\AA] &  --2.9404 $\pm$  0.5046 &  1.4908 $\pm$  0.6961 &     --0.4374 $\pm$     0.3443 &     0.0800 $\pm$     0.0588 & 24.4 & [  --2.12,  2.75 ] \\
  C$_{2}$4668~[\AA] &  --1.8760 $\pm$  0.1041 &  1.3531 $\pm$  0.1605 &     --0.4178 $\pm$     0.0986 &     0.0467 $\pm$     0.0223 & 87.3 & [  --2.12,  3.38 ] \\
\enddata
\tablenotetext{a}{BIC=Bayesian information criterion}
\label{tab:coeff3}
\end{deluxetable}

\clearpage
\begin{figure}
 \begin{center}
\includegraphics[scale=0.85]{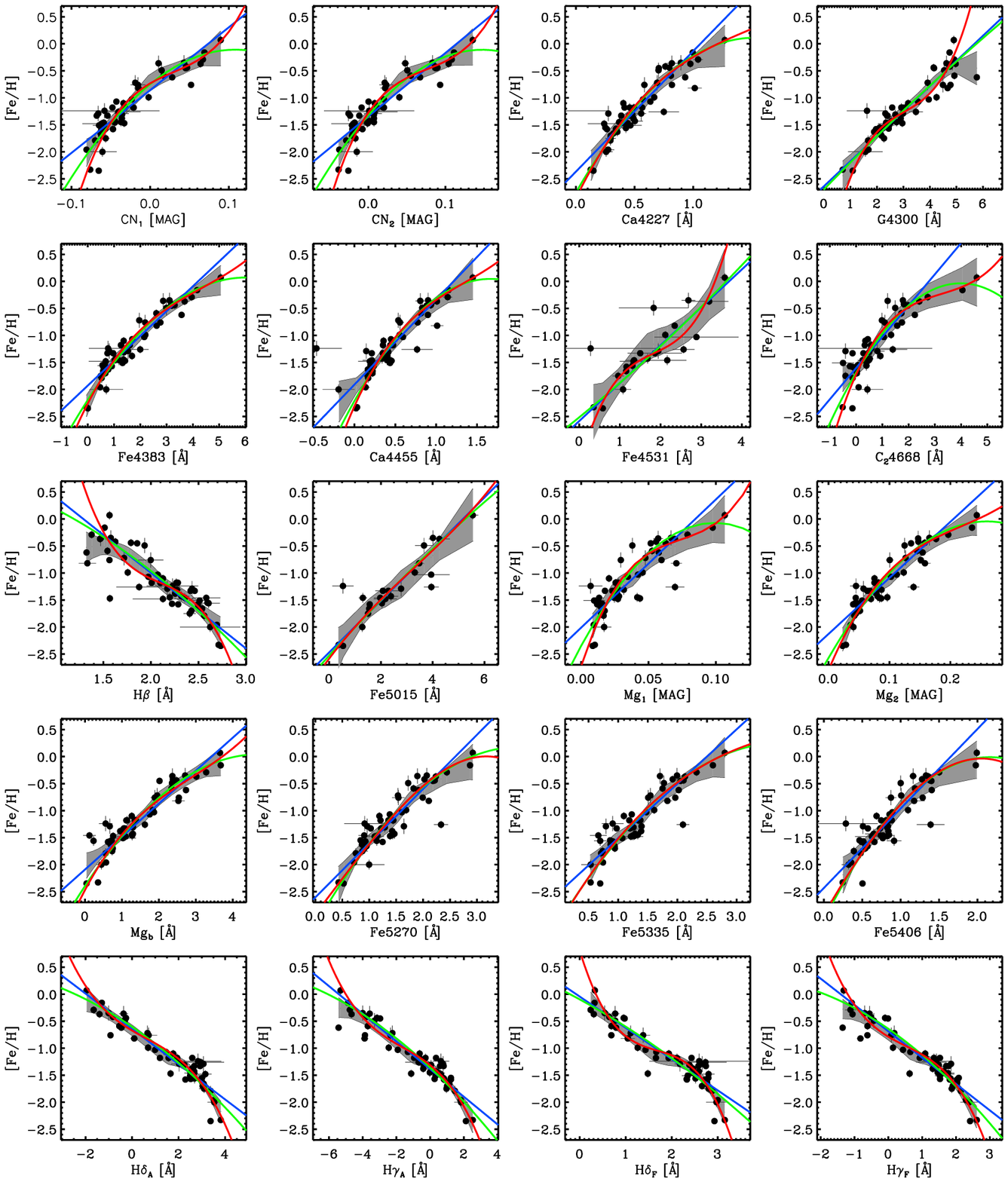}
\caption{The LIS index--metallicity relations. 
The blue, green, and red solid lines represent the first-, second-, and third-order orthogonal least-squares fit, respectively. The gray-shaded regions indicate the 95\% confidence bands based on the LOESS regression, which show the underlying trend of the data.}
  \label{fig:LISmet}
 \end{center} 
\end{figure}

\clearpage
\begin{deluxetable}{lrrrr}
\tablecolumns{5}
\tabletypesize{\small}
\rotate
\tablecaption{Coefficients of the LIS index versus metallicity relations. The coefficients were obtained by the orthogonal distance least-squares fit of a linear function $\rm[Fe/H]\it=a_{0}+a_{1} \times (index)$ to 53 GGC data.}
\tablewidth{0pt}
\tablehead{\colhead{Index} & \colhead{$a_{0}$} & \colhead{$a_{1}$} & \colhead{BIC\tablenotemark{a}} & \colhead{Valid Range}}
\startdata
     CN$_{1}$~[mag] &~~~  --0.8572 $\pm$  0.0293 ~~~&~~~ 11.6263 $\pm$  0.5863 ~~~& 38.1 & [  --1.4887,  2.5112 ] \\
     CN$_{2}$~[mag] &~~~  --1.3408 $\pm$  0.0290 ~~~&~~~ 11.7227 $\pm$  0.6203 ~~~& 42.8 & [  --2.0814,  2.5112 ] \\
       Ca4227~[\AA] &~~~  --2.3670 $\pm$  0.0660 ~~~&~~~  2.2327 $\pm$  0.1102 ~~~& 46.3 & [  --2.12,  2.51 ] \\
        G4300~[\AA] &~~~  --2.6373 $\pm$  0.0689 ~~~&~~~  0.4643 $\pm$  0.0205 ~~~& 29.6 & [  --1.49,  2.62 ] \\
       Fe4383~[\AA] &~~~  --1.9284 $\pm$  0.0398 ~~~&~~~  0.4596 $\pm$  0.0199 ~~~& 12.4 & [  --2.08,  2.62 ] \\
       Ca4455~[\AA] &~~~  --1.9097 $\pm$  0.0404 ~~~&~~~  1.5739 $\pm$  0.0684 ~~~& 23.0 & [  --2.12,  2.62 ] \\
       Fe4531~[\AA] &~~~  --2.5856 $\pm$  0.1207 ~~~&~~~  0.7018 $\pm$  0.0608 ~~~& 25.8 & [  --2.12,  2.78 ] \\
  C$_{2}$4668~[\AA] &~~~  --1.6046 $\pm$  0.0467 ~~~&~~~  0.5787 $\pm$  0.0395 ~~~& 57.0 & [  --2.12,  2.78 ] \\
\enddata
\tablenotetext{a}{BIC=Bayesian information criterion}
\label{tab:LIScoeff1}
\end{deluxetable}

\clearpage
\begin{deluxetable}{lrrrrr}
\tablecolumns{6}
\tabletypesize{\small}
\rotate
\tablecaption{The same as Table 9, but using a quadratic polynomial $\rm[Fe/H]\it=a_{0}+a_{1} \times (index) + a_{2} \times (index)^2$ to the 53 GGC data.}
\tablewidth{0pt}
\tablehead{\colhead{Index} & \colhead{$a_{0}$} & \colhead{$a_{1}$} & \colhead{$a_{2}$} & \colhead{BIC\tablenotemark{a}} & \colhead{Valid Range}}
\startdata
     CN$_{1}$~[mag] &  --0.7465 $\pm$  0.0408 & 11.8023 $\pm$  1.0299 &   --54.4875 $\pm$ 13.3298 & 33.7 & [  --1.4887,  2.5112 ] \\
     CN$_{2}$~[mag] &  --1.3196 $\pm$  0.0392 & 16.3757 $\pm$  0.9635 &   --55.4078 $\pm$ 13.4646 & 39.8 & [  --2.0814,  2.5112 ] \\
       Ca4227~[\AA] &  --2.7543 $\pm$  0.1225 &  3.7625 $\pm$  0.3828 &    --1.2343 $\pm$  0.3356 & 41.0 & [  --2.12,  2.51 ] \\
        G4300~[\AA] &  --2.6957 $\pm$  0.2014 &  0.5063 $\pm$  0.1202 &    --0.0066 $\pm$  0.0187 & 32.7 & [  --1.49,  2.62 ] \\
       Fe4383~[\AA] &  --2.1948 $\pm$  0.0608 &  0.7540 $\pm$  0.0505 &    --0.0627 $\pm$  0.0133 & 18.9 & [  --2.08,  2.62 ] \\
       Ca4455~[\AA] &  --2.2119 $\pm$  0.0688 &  2.7134 $\pm$  0.2019 &    --0.8150 $\pm$  0.1859 & 27.3 & [  --2.12,  2.62 ] \\
       Fe4531~[\AA] &  --2.5163 $\pm$  0.2575 &  0.6147 $\pm$  0.2505 &    0.0224 $\pm$  0.0671 & 28.9 & [  --2.12,  2.78 ] \\
  C$_{2}$4668~[\AA] &  --1.6931 $\pm$  0.0517 &  0.8317 $\pm$  0.0511 &    --0.1043 $\pm$  0.0180 & 53.6 & [  --2.12,  2.78 ] \\
\enddata
\tablenotetext{a}{BIC=Bayesian information criterion}
\label{tab:LIScoeff2}
\end{deluxetable}

\clearpage
\begin{deluxetable}{lrrrrrr}
\tablecolumns{7}
\tabletypesize{\small}
\rotate
\tablecaption{The same as Table 9, but using a cubic polynomial $\rm[Fe/H]\it=a_{0}+a_{1} \times (index) + a_{2} \times (index)^2 + a_{3} \times (index)^3$ to the 53 GGC data.}
\tablewidth{0pt}
\tablehead{\colhead{Index} & \colhead{$a_{0}$} & \colhead{$a_{1}$} & \colhead{$a_{2}$} & \colhead{$a_{3}$} & \colhead{BIC\tablenotemark{a}} & \colhead{Valid Range}}
\startdata
     CN$_{1}$~[mag] &  --0.7321 $\pm$  0.0566 &  8.5305 $\pm$  1.9158 &    --76.9963 $\pm$    36.2398 &   871.5720 $\pm$   328.5897 & 51.4 & [  --1.4887,  2.5112 ] \\
     CN$_{2}$~[mag] &  --1.2693 $\pm$  0.0504 & 20.4390 $\pm$  1.5696 &   --196.3766 $\pm$    30.7768 &   865.7252 $\pm$   300.5097 & 50.7 & [  --2.0814,  2.5112 ] \\
       Ca4227~[\AA] &  --2.8616 $\pm$  0.2526 &  4.4588 $\pm$  1.1041 &     --2.3699 $\pm$     1.7731 &     0.5285 $\pm$     1.0364 & 52.2 & [  --2.12,  2.51 ] \\
        G4300~[\AA] &  --4.2812 $\pm$  0.6967 &  2.4309 $\pm$  0.6108 &     --0.7105 $\pm$     0.1865 &     0.0783 $\pm$     0.0194 & 54.3 & [  --1.49,  2.62 ] \\
       Fe4383~[\AA] &  --2.2594 $\pm$  0.1013 &  0.9426 $\pm$  0.1160 &     --0.1558 $\pm$     0.0517 &     0.0120 $\pm$     0.0090 & 26.6 & [  --2.08,  2.62 ] \\
       Ca4455~[\AA] &  --2.3102 $\pm$  0.1116 &  3.3596 $\pm$  0.4524 &     --1.8492 $\pm$     0.7002 &     0.4571 $\pm$     0.4229 & 31.2 & [  --2.12,  2.62 ] \\
       Fe4531~[\AA] &  --3.4107 $\pm$  0.5701 &  2.6925 $\pm$  0.7989 &     --1.2758 $\pm$     0.4152 &     0.2322 $\pm$     0.0754 & 33.9 & [  --2.12,  2.78 ] \\
  C$_{2}$4668~[\AA] &  --1.7261 $\pm$  0.0739 &  1.0741 $\pm$  0.1088 &     --0.2826 $\pm$     0.0682 &     0.0288 $\pm$     0.0159 & 62.6 & [  --2.12,  2.78 ] \\
\enddata
\tablenotetext{a}{BIC=Bayesian information criterion}
\label{tab:LIScoeff3}
\end{deluxetable}

\end{document}